\documentclass[runningheads]{llncs}
\usepackage{bbding}
\usepackage[misc]{ifsym}
\usepackage{amsmath,amsfonts}
\usepackage{algorithmic}
\usepackage{hyperref}

\usepackage{graphicx}
\usepackage{textcomp}
\usepackage{subfigure}
\usepackage{booktabs}
\usepackage{multirow}
\usepackage{xspace}
\usepackage{xcolor}
\usepackage{listings}
\usepackage{balance}
\usepackage{diagbox}

\hypersetup{hidelinks}

\newcommand{\ie}{\emph{i.e.,}\xspace}
\newcommand{\eg}{\emph{e.g.,}\xspace}
\newcommand{\resp}{\emph{resp.,}\xspace}

\newcommand{\eat}[1]{}

\def\BibTeX{{\rm B\kern-.05em{\sc i\kern-.025em b}\kern-.08em
    T\kern-.1667em\lower.7ex\hbox{E}\kern-.125emX}}

\begin{document}

\title{Enc$^2$DB: A Hybrid and Adaptive Encrypted Query Processing Framework
}
%
%
\author{Hui Li\inst{1,2(\textrm{\Letter})} \and
Jingwen Shi\inst{1}\and
Qi Tian\inst{1} \and
Zheng Li\inst{1} \and
Yan Fu\inst{1} \and
Bingqing Shen\inst{3} \and
Yaofeng Tu\inst{4}}
\authorrunning{H.Li et al.}
%
\institute{ Xidian University, Xi'an,China \\
\email{hli@xidian.edu.cn, \{jingwen.shi,tianqi,lizhen,yanfu\}@stu.xidian.edu.cn}\and
Shanghai Yunxi Technology Co., Ltd., Shanghai, China\\
\email{lihui22@inspur.com}\and
Shanghai International Studies University, Shanghai, China\\
\email{bqshen@shisu.edu.cn}\and
ZTE Corporation, Nanjing, China \\
\email{tu.yaofeng@zte.com.cn}}                                                                                                                                                                                                                          
\maketitle              
\begin{abstract}
As cloud computing gains traction, data owners are outsourcing their data to cloud service providers (CSPs) for Database Service (DBaaS), bringing in a deviation of data ownership and usage, and intensifying privacy concerns, especially with potential breaches by hackers or CSP insiders.
To address that, encrypted database services propose encrypting every tuple and query statement before submitting to the CSP, ensuring data confidentiality when the CSP is honest-but-curious, or even compromised. Existing solutions either employ property preserving cryptography schemes, which can perform certain operations over ciphertext without decrypting the data over the CSP, or utilize trusted execution environment (TEE) to safeguard data and computations from the CSP. Based on these efforts, we introduce Enc2DB, a novel secure database system, following a hybrid strategy on PostgreSQL and openGauss. We present a micro-benchmarking test and self-adaptive mode switch strategy that can dynamically choose the best execution path (cryptography or TEE) to answer a given query. Besides, we also design and implement a ciphertext index compatible with native cost model and query optimizers to accelerate query processing. Empirical study over TPC-C test justifies that Enc2DB outperforms pure TEE and cryptography solutions, and our ciphertext index implementation also outperforms the state-of-the-art cryptographic-based system.

\end{abstract}
\section{Introduction}
Cloud computing has become an essential infrastructure for efficient data management systems. Migrating workloads to the cloud brings many advantages, such as lower cost, higher flexibility, greater scalability, higher reliability and so on. More and more enterprises and institutions are tending to rely on third-party database service providers for storing and managing their data services.

However, while storage and computing in the cloud bring great convenience, the deviation of data ownership and usage also brings privacy risks that can not be ignored. When users outsource data to cloud service providers (CSP), they lose physical control over data. The security and privacy of data depend on the security policy provided by the CSP. If the security policies are breached by external hackers or even rogue employees from the CSP itself, users' sensitive data can be leaked, seriously compromising data security and privacy.

To ensure confidentiality, data can be kept encrypted over the CSP. Traditional data encryption~\cite{ArvindArasu2013OrthogonalSW,monomi} preserves confidentiality of data at rest. Securing data at rest allows users to utilize cloud storage without exposing sensitive information, but it prevents users from performing SQL queries effectively or efficiently.

A possible solution to this problem is homomorphic encryption, which can provide data operability while preserving data confidentiality. The fully homomorphic encryption (FHE)~\cite{gentry2009fully} scheme allows algebraic computations over encrypted data, but it suffers from high overhead for complex analytical queries. Partially homomorphic encryption (PHE)~\cite{rivest1978data} allows computations over encrypted data with respect to some specific operations and have practical performance. Property-Preserving
Encryption (PPE)~\cite{pandey2012property} can preserve some attributes of plaintext data, such as comparisons. 

Another alternative is to utilize trusted execution environment to ensure the confidentiality of data. The trusted execution environment (TEE)~\cite{sabt2015trusted} constructs a secure area in the CPU (called an enclave) through software and hardware methods, ensuring that the programs and data loaded inside are protected in terms of confidentiality and integrity. Therefore, the encrypted data can be decrypted and calculated in TEE, such that the OS knows nothing about the content inside. Due to recent advancements in TEE, many enclave-based encrypted databases and storage systems have emerged~\cite{antonopoulos2020azure,bailleu2019speicher,eskandarian2017oblidb,kim2019shieldstore,mishra2018oblix,priebe2018enclavedb,vinayagamurthy2019stealthdb,zheng2017opaque,sun2021building}. For example, EnclaveDB~\cite{priebe2018enclavedb} protects the confidentiality and integrity of data and queries by placing sensitive data (tables, indexes and other metadata) in enclaves protected by trusted hardware (such as SGX). 

In this paper, we propose a hybrid and adaptive encryption query processing solution, namely Enc$^2$DB, implemented in our openGauss, an open-source database proposed by Huawei, as well as in PostgreSQL. Based on the cryptography technologies such as symmetric PHE, order preserving encryption (OPE), and AES we establish the ciphertext storage model of relational database and the corresponding query processing framework. Enc$^2$DB proposes a hybrid solution make use of both software (cryptography) and hardware (TEE) to improve the efficiency of ciphertext data query, and realizes the fully encrypted storage and execution of query workload, as well as the transparent processing of user-side query requests. Our main contributions are as follows:
\begin{itemize}
\item We use full ciphertext-based storage, support user defined column-level choice of plaintext or ciphertext state.
\item We propose a ciphertext-aware indexing mechanism over ORE, to further improve query efficiency.
\item We combine SGX with software-only solutions and propose a hybrid self-adaptive strategy towards range queries.
\end{itemize}

In the rest of the paper, we first present background information and related work in \autoref{sec2} and then explores system architecture in \autoref{sec3}, as well as the implementation details of software and TEE-enabled modes. In \autoref{sec4}, we introduce a series of optimizations. Empirical study is provided in \autoref{sec5}, and we conclude in \autoref{sec6}.

\section{BACKGROUND AND RELATED WORK}\label{sec2}
\subsection{Encryption Algorithm}
\subsubsection{Homomorphic Encryption}
Homomorphic encryption is proposed for mathematical calculation of ciphertext without obtaining the key. According to the types of ciphertext calculation, it can be divided into FHE and PHE. FHE allows any algebraic operation on encrypted data, and the result obtained after decryption preserves the calculation result.

In 2009, Gentry~\cite{gentry2009fully} implemented the first fully homologous encryption scheme. Although over the years, the algorithm of FHE has experienced significant improvement in the aspect of efficiency, it is impractical for real-world usage. In comparison, PHE only supports limited types of ciphertext calculation, but provide practical efficiency. PHE is mainly divided into additive homomorphic encryption (AHE) and multiplicative homomorphic encryption (MHE). Examples of AHE include Paillier~\cite{paillier1999public}, benaloh~\cite{fousse2011benaloh}, Okamoto–Uchiyama~\cite{okamoto1998epoc}, Naccache stern~\cite{naccache1998new}, Damg{\aa}rd-Jurik\cite{damgaard2001generalisation},etc. Examples of MHE include ElGamal~\cite{elgamal1985public} and RSA~\cite{rivest1978method}. At present, we adopt the state-of-the-art symmetric AHE and MHE solution, namely SAHE and SMHE proposed in Symmetria~\cite{savvides2020efficient}. They retain the full range of homomorphic operations that asymmetric schemes support while offering the same level of security (semantic security, (IND-CPA)).

\subsubsection{Property Preserving Encryption}
For operations such as range queries involving comparisons that HE cannot support, we turn to the Property Preserving Encryption (PPE), which has been used in systems such as CryptDB~\cite{popa2012cryptdb}, Monomi~\cite{monomi} and Seabed~\cite{papadimitriou2016big}.

The ciphertext of the PPE scheme retains some attributes of the plaintext. For the same plaintext data, the ciphertext encrypted by deterministic encryption (DET) is also the same. There are other schemes that can preserve the order relationship of the underlying plaintext values, which is called Order Preserving Encryption (OPE)~\cite{agrawal2004order}. The relative order of the plaintext values can be directly obtained by comparing the encrypted values. Boldyreava~\cite{boldyreva2011order} proposes an attribute-preserving encryption scheme, but it's a deterministic scheme that preserves frequency information of plaintext data. At present, the Order-Revealing Encryption scheme (ORE)~\cite{boneh2015semantically,chenette2016practical,1555162} and its branch scheme are non-deterministic solutions, which can reveal less information under the premise of guaranteeing higher operation efficiency, thus providing better security than OPE. 

\subsection{Trusted Execution Environment}
Trusted Execution Environment(TEE)~\cite{sabt2015trusted,DBLP:journals/pvldb/XiaZZZCEFFKF22} builds a secure area in the central processing unit through software and hardware methods to ensure that the programs and data loaded in it are protected in terms of confidentiality and integrity. The principle of TEE is to divide the hardware and software resources of the system into two execution environments-trusted execution environment and common execution environment. The two environments are securely isolated, with independent internal data paths and storage space required for computing. Applications in a common execution environment cannot access the TEE. Even within the TEE, multiple applications run independently and cannot access each other without authorization. All major CPU vendors have rolled out their TEE (\eg ARM TrustZone, Intel SGX, and AMD SEV) to provide a secure execution environment, commonly referred to as an enclave~\cite{intel-sgx,ARM-TrustZone,AMD-SEV}.

Encrypted database systems can employ TEE to preserve the confidentiality in query processing, by decrypting the ciphertext and performing complex calculations over plaintext within TEE.  SGX is a representative TEE framework proposed by Intel, which encapsulates the security operations of legitimate software in an enclave to protect it from malware attacks. That is, even the OS or the VMM (Hypervisor) cannot access or affect the code and data inside the enclave.

EnclaveDB~\cite{priebe2018enclavedb} relies on SGX and processes encrypted queries  in an enclave, which allows only pre-compiled queries and assumes that all data can fit in the memory. Always encrypted Azure database~\cite{antonopoulos2020azure} uses a few enclave-based defined function for computation over ciphertext. However, such non-intrusive design leads to possible information leakage and performance degradation. FE-in-GaussDB~\cite{zhu2021full} combines encryption algorithms with TEE to securely perform various operations on ciphertext data, including matching, comparison and etc.  

\section{system architecture}\label{sec3}
In this section, we introduce the basic structure of Enc$^2$DB (Encrypted Database with Enclave), which provides two deployment modes, \ie software-only and TEE-enabled. To start with, we first propose the software-only mode, namely EncDB (Encrypted Database). After that, we discuss how TEE can co-operate with EncDB by enabling complex tasks to run inside the Enclave, which leads to the second mode, Enc$^2$DB (detailed in \autoref{hardware-based approach}).


\begin{figure}
\includegraphics[width=\textwidth]{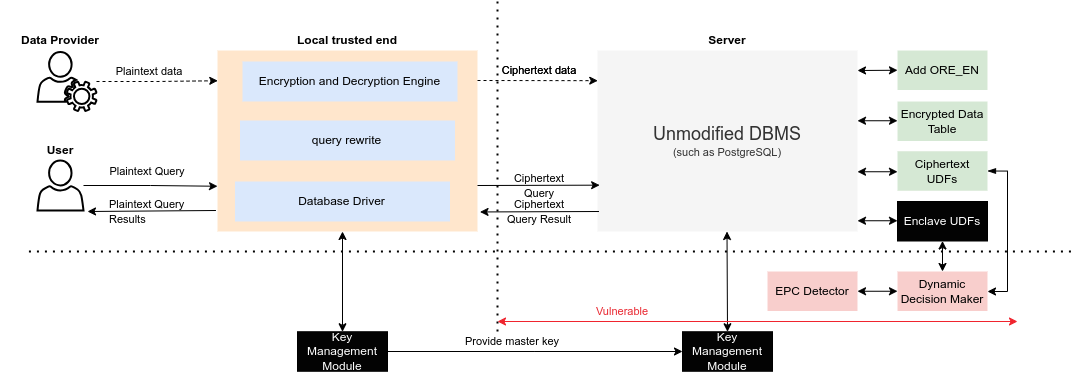}
\caption{System Architecture with components differentiated by mode: TEE-enabled (black and red) and Software-based (green).} \label{fig.4.1}
\end{figure}

The complete system is divided into two parts: the application server (trusted client) and the untrusted server. Its architecture is shown in Fig.~\ref{fig.4.1}, which includes the following components deployed over both the client and the cloud.

\noindent\textbf{Encryption and Decryption Engine:} holds the encryption key, encrypts plaintext data, and decrypts the result received from the cloud. The encryption module also generates and stores meta-encryption data, \ie the mapping of plaintext operators to the relevant ciphertext ones, the supported encryption schemes towards different data types in the database.

\noindent\textbf{Query rewriter:} convert plaintext queries into the ones in ciphertext, which are sent through the network to the server.

\noindent\textbf{Database Driver: } PostgreSQL supports multiple database application programming interfaces (such as libpq, JDBC, ODBC, etc.) to submit query requests. The application driver helps the client to send the rewritten encrypted SQL to the server and receive the encrypted result from the cloud.

\noindent\textbf{Server-side Computing Services: } Leverage a set of user-defined functions (UDFs) to perform operations on encrypted data.

The pipeline for answering a SQL query under the architecture is as follows: The data provider uploads plaintext data, and the client encrypts each column under one or more encryption schemes according to the expected operations, replacing the original column name with a random string for anonymization. The user enters a plaintext query, which is rewritten and sent to the server in the form of a ciphertext counterpart. The server implements the relevant operation logic for ciphertext calculation in advance, performs query and operation on the encrypted SQL, and returns the obtained encrypted result to the client. The client decrypts the received ciphertext to obtain the plaintext query result. The whole process is completely transparent to the user, and the server is unaware about the content of either the query or the result.

\subsection{Software-based mode, EncDB}
In our software-only implementation mode, a column in a table is to be stored in ciphertext form under one (or more) of AHE, MHE, ORE, and AES. Given a SQL statement, the column names and operands are respectively replaced using desensitized names and ciphertext under a predefined encryption logic.  This process takes place at the user's local trusted end and is deployed by a small connection pool, which has many advantages such as transparency to the upper layers, no client awareness, and easy migration. After receiving the encrypted query statements, the server will perform the relevant ciphertext calculation according to the pre-configured user-defined functions, and finally returns the encrypted query results to the user, and the local trusted end will decrypt and obtain the plaintext results, which is completely transparent to the application. 

\noindent\textbf{Advantages}. A major benefit of the software-only implementation mode is that it exert no hardware requirement, and it also has less deployment costs.

\noindent\textbf{Disadvantages}. The obvious disadvantage of the software-only mode is the large redundancy of data, as a data column is copied and encrypted into many ciphertext columns under different encryption schemes, resulting in a huge space and time overhead. 
In addition, the query work is done using UDFs with a huge computational overhead. 
Besides, some complex expressions are hard to compute under this mode. For example, none of current encryption schemes can support both addition/multiplication homomorphic and comparison in ciphertext space. For queries containing both operations over the same operands, another round of interaction between the client and server is extraly required. This solution introduces network communication latency, which is unacceptable for a real-time online transaction processing system. 

The software-only mode is a major research direction in  cryptographic database systems \cite{hacigumucs2002executing,poddar2016arx,popa2012cryptdb,monomi,sdb}. The EncDB in our system is an advanced solution in this trend that absorbs many of the advantages of the previous works, especially those symmetric encryption schemes. Based on this implementation, we further introduce TEE and propose a hybrid solution that benefit from both software and hardware security support. Besides, we also present a self-adaptive strategy to enable both techs cooperate at runtime, which results in a large improvement in the overall performance of the system. 

\subsection{TEE-enabled mode, Enc$^2$DB}\label{hardware-based approach}
\begin{figure}
\includegraphics[width=\textwidth]{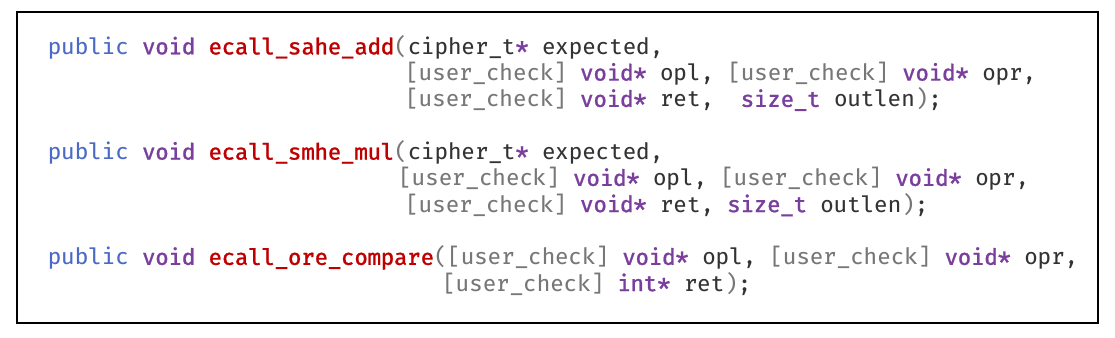}
\caption{EDL Definition of UDFs} \label{edl}
\end{figure}

The TEE-enabled mode aims to improve the efficiency and address the unsupported operations under software-only mode. In the TEE-enabled mode implementation, UDFs are packaged as independent trusted bridging functions, which are used to enter secure memory, \eg Enclave, decrypt encrypted data, and encrypt the results according to the required encryption mode after completing the corresponding computation task. Finally, the computed results are returned to the database system, which also maintains the high security of the data to the host. Since the computation is done in the secure area, we never need to worry about homomorphic or property preserving functionality in the encryption scheme. Instead, any symmetric encryption can be considered, as long as it is enough efficient and secure. In addition, comparing to the presence of various encrypted copies (each corresponds to a specific encryption scheme) in software-only mode, only one encrypted copy is stored at the server side. Suppose Intel SGX is used in Fig.~\ref{fig.4.1}, trusted bridging functions are declared in SGX via the Enclave Definition Language (EDL), and the trusted settlement interface for some UDFs is shown in Fig.~\ref{edl}. The first parameter is the encryption mode of the result to be calculated. The next two parameters are the left and right operands computed for this ciphertext, since SGX can directly access the address space of the host process, there is no need for a memory copy. The fourth parameter is the ciphertext calculation result. Since the result is generated in secure memory, a memory copy will occur here, aiming to move data from secure area to the unsecure memory space.



\begin{figure}
\includegraphics[width=\textwidth]{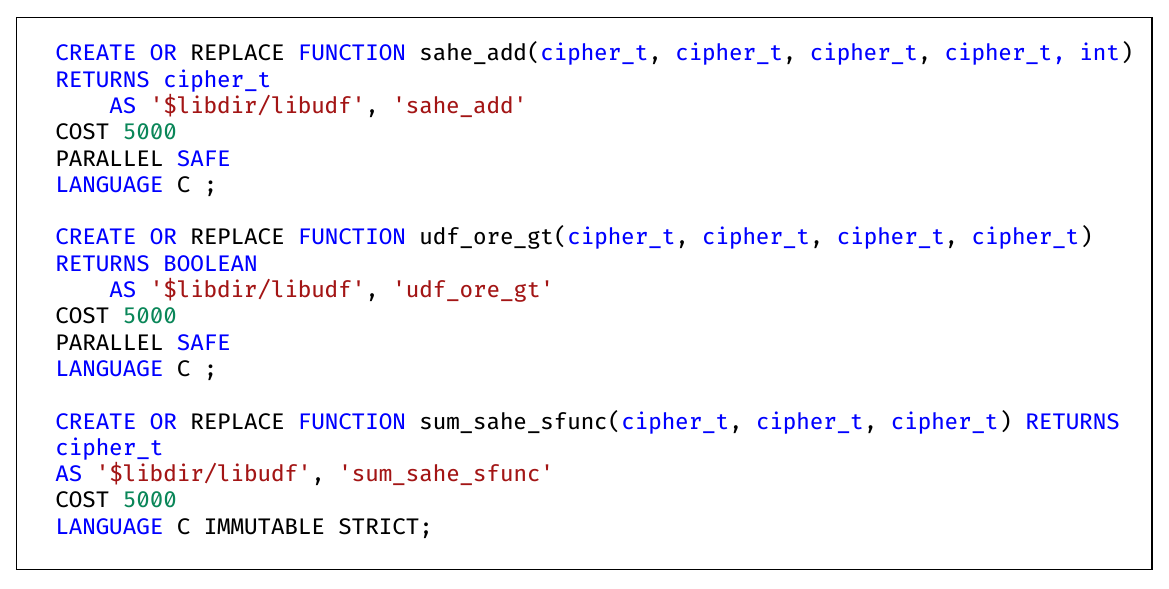}
\caption{Database System Partial UDF Definitions} \label{udfs}
\end{figure}

The database-level UDF definition is shown in Fig.~\ref{udfs}. The first two parameters it accepts are operand in AES ciphertext and the last two are ciphertext under homomorphic encryption (\eg AHE, MHE), which is employed to support self-adpative hybrid query processing in \autoref{ssec:co}. In addition, as cryptographic UDF calculations are more expensive than plaintext ones, a nifty query plan should perform these ciphertext UDF calculations as late as possible. In light of that, we inject and assign in the cost model these UDFs a high enough cost.

\subsubsection{Key Transfer Protocol and Management}
\begin{figure}
\includegraphics[width=\textwidth]{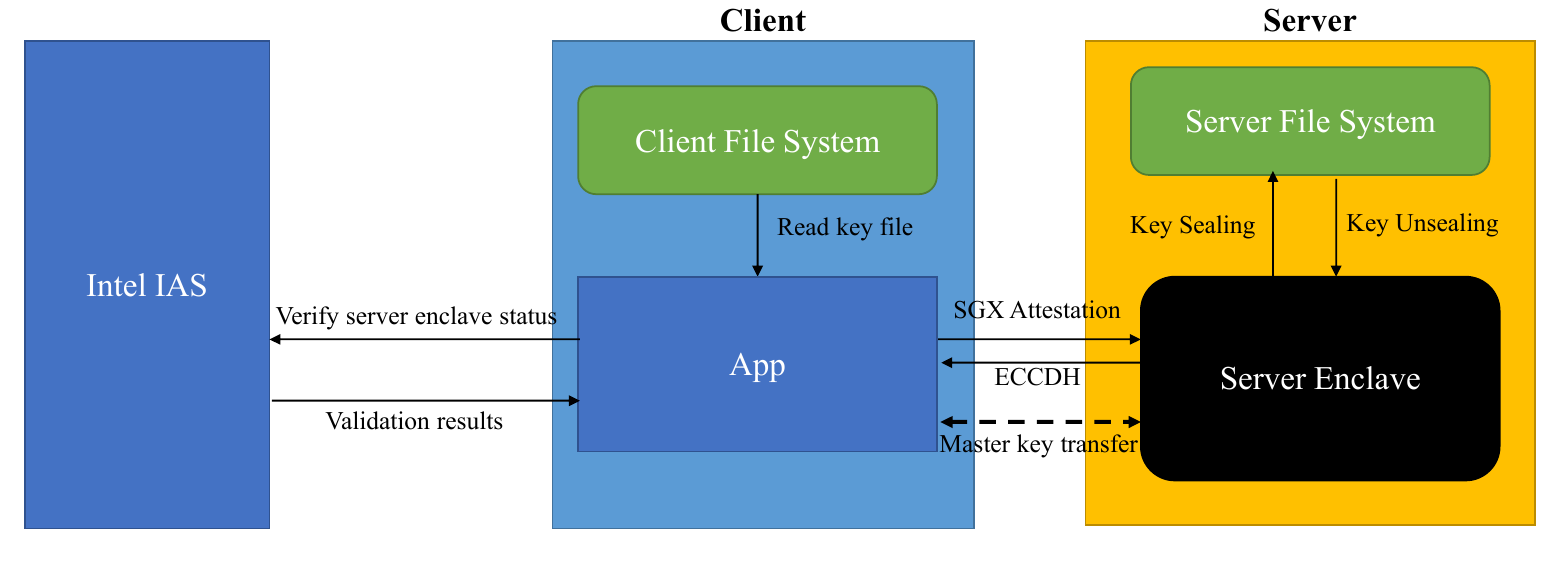}
\caption{Key Transfer Based on Remote Attestation} \label{fig.4.5}
\end{figure}

Since all secure calculations in Enclave are conducted on plaintext data, keys that exist on the local trusted side must be securely transferred to the Enclave instances running in the cloud over untrusted channels. The solution adopted here is achieved by using SGX remote authentication technology. During the deployment phase of a secret database, the remote and local trusted ends establish a secure communication channel through remote authentication. Afterwards, the master key is transferred to Enclave, where the encryption keys are derived for each encrypted column internally through the key export algorithm HKDF~\cite{cryptoeprint:2010:264}, and save them to the file device of the host machine through SGX sealing technology for later use when the restart is applied. The overall flow of key transfer is shown in Fig.~\ref{fig.4.5}.

At the beginning of the key transfer protocol, the client starts the remote authentication process by sending an initial message (\ie $init$) to the server. After receiving the $init$ message, the server will generate $msg0$, including $EPID$ of the server. When the client receives $msg0$, $epid$ will be registered locally, indicating that a link is being established with this server, and returning the message of successful registration, or failing and terminating the following process if it has already registered before. When the server receives a message confirming the success of $msg0$, it generates $msg1$ in the Enclave security zone, the primary data field is the public key portion of the Elliptic Curve Key Exchange Algorithm (ECCDH), and sends $msg1$ to the client. The client receives $msg1$, first gets the signature revocation list through Intel IAS, then generates the ECCDH public key $Gb$ locally, and calculates the shared key $DH\_KEY$ based on the derivation of the ECCDH public key $Ga$ on the server, and then derives $MK$, $SK$ and $VK$ from $DH\_KEY$ in turn to meet future communication needs. Afterwards, client public key $Gb$, CMAC authentication code $cmac$ are generated with $MK$. Use predefined server-side public key signatures of $Ga$ and $Gb$ are employed to prevent man-in-the-middle attacks. Client assembles all the above information into $msg2$ and sends it to the server. When the server receives $msg2$, it checks each data in the security area and generates Enclave's $quote$ to be sent to the client in $msg3$. $quote$ is the key proof of Enclave's validity. When a client receives $msg3$, $Ga$ in $msg3$ is first validated against the client's own copy, then $MK$ is used to verify that $cmac$ is correct or not, and then $quote$ is formatted for $Ga+Gb+VK$. Upon successful verification, $quote$ is sent to Intel IAS for authentication to confirm that the other party (at server side) is an application running in a real SGX environment. The verification results are assembled to generate attestation messages that are sent to the server to notify it about the validation and establish a secure communication channel. Attestation messages can carry custom information, encrypted with $SK$ and stored in the payload field, encrypted with AES-GCM, and checkcodes stored in the payload tag field. At this point, the key transfer protocol is complete and subsequent communications can be made through the shared key of $SK$ via secret messages. 

\vspace{-1ex}\subsection{Self-adaptive switch between TEE-enabled and software modes}\label{ssec:co}
Using trusted bridging functions to wrap UDFs can support arbitrary type of query over ciphertexts, but trusted hardware also suffers from a series of limitations, \ie insufficient safe memory space for SGX~\cite{bailleu2019speicher}~\cite{arnautov2016scone}~\cite{orenbach2017eleos}, which can lead to severe page replacement in concurrent transaction processing scenarios, even worse than that of software-only implementations. Therefore, outsourcing all queries to trusted hardware cannot fit for all practical scenarios.
Driven by that, we select to mitigate the problem by processing the encrypted query using both software and TEE-enabled modes together. Depending on whether the processing is dynamically switched between both modes at runtime or not, we propose two strategies, namely static switch and self-adaptive switch.

\subsubsection{Static mode switch}
Specifically, in static switch strategy, for those operations efficiently addressed by software mode, we select not to rely on TEE. To this end, we conduct both theoretical and empirical study over the encryption schemes to tell which can be efficiently addressed in software-only mode. Fig.~\ref{5co_pie} is a comparison of the time consumed for both AHE and ORE between both modes over TPC-C workload. Obviously, ORE has the largest proportion, which is consistent with the fact that ORE has greater space and time complexity than other encryption schemes. Intuitively, ORE-related operations is the first target we must kill (\ie transfer the task to TEE) in software-only mode. 

All operations involving OREs can be replaced with AES with the help of TEE to save time and space. Assuming that ORE operations are statically replaced using AES and supported in TEE, a SQL query shall replace the corresponding columns involved in all comparison predicates using AES columns, instead of ORE ones. On the server side, UDFs involving comparison over columns, are all moved to Enclave, which shall decrypt the incoming AES ciphertext, perform comparison over plaintext, and return the corresponding comparison results. This static configuration for performing a fixed type of predicates on ciphertext operands, \ie comparison, over TEE is referred to as a static switching mechanism.

If all predicates of ciphertext operands are moved and fully relied on TEE, static switch becomes the TEE-enabled mode shown in \autoref{hardware-based approach}. This simple replacement of an encrypted column to reduce space-time overhead does improve the overall performance of an encrypted database system. If all types of predicates over ciphertext operands are handed over to SGX, it works well in small-scale or low-concurrency scenarios. However, in high concurrency scenarios, SGX's secure memory space will soon be full, resulting in severe paging. According to the previous analysis, if page fault occurs, all operations will take almost twice time as long as those without page fault. 

\begin{figure}
    \centering
    \subfigure[Binary Search in SGX]{\label{2EPCbsearch} \includegraphics[width=.4\linewidth]{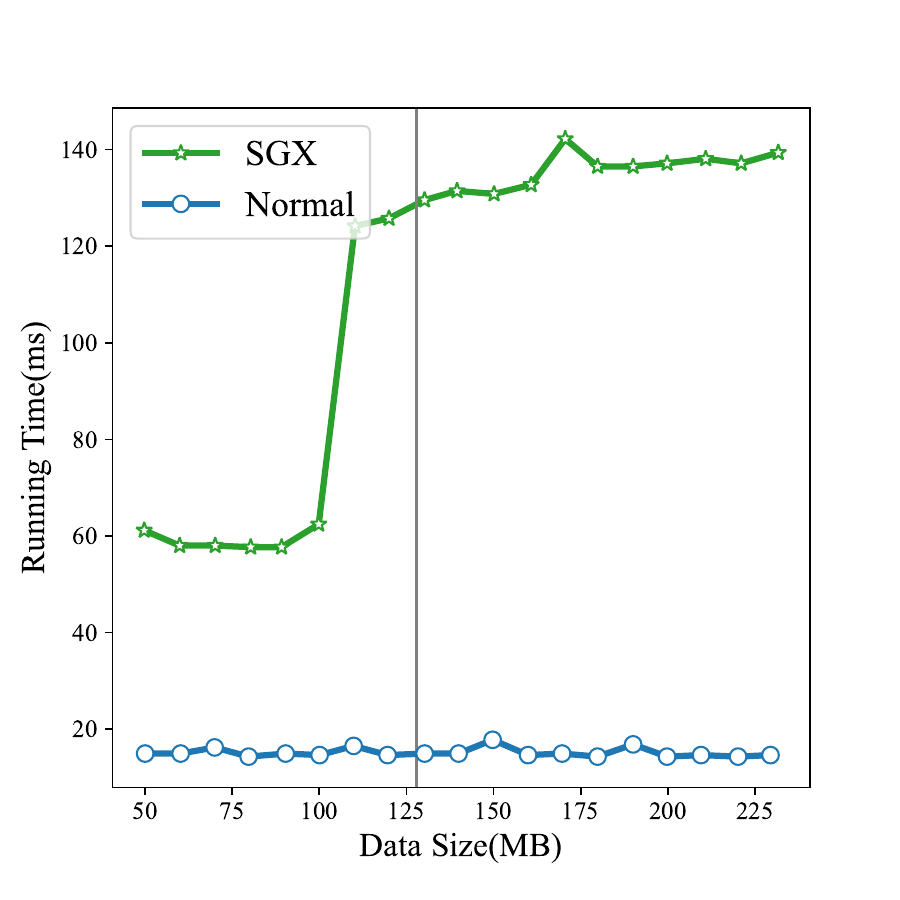}}
    \subfigure[Quick Sort in SGX]{\label{2EPCqsort} \includegraphics[width=.4\linewidth]{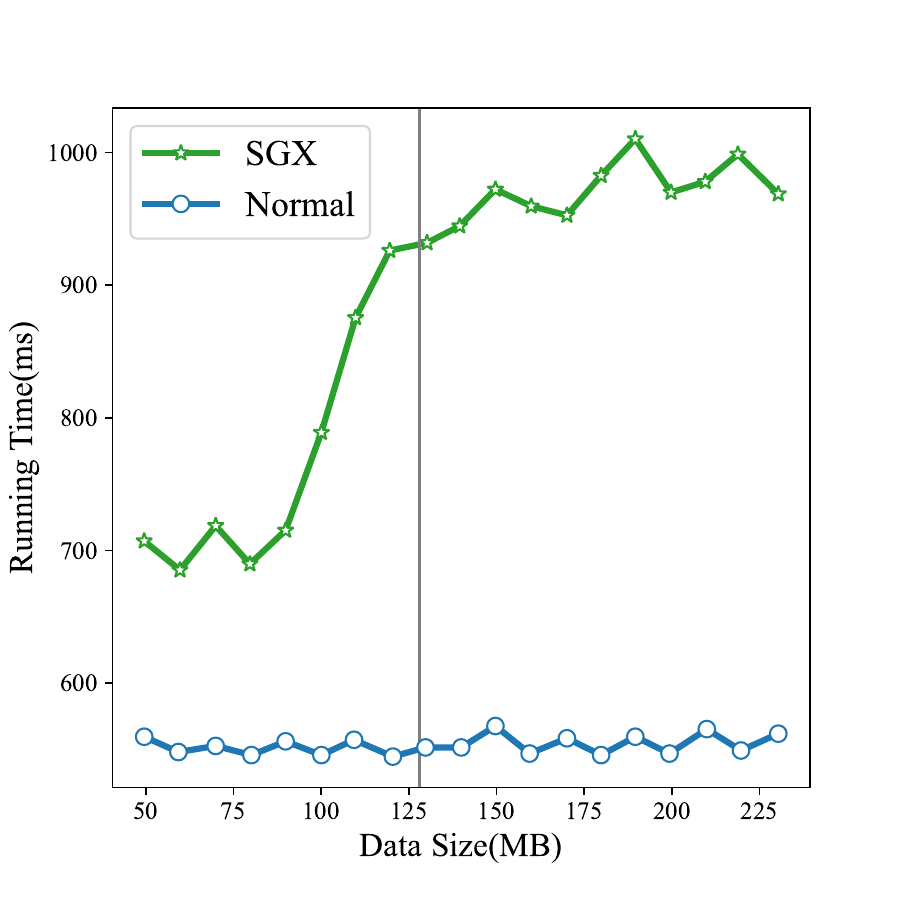}}
    \vspace{-1ex}\caption{Execution efficiency of classical algorithms in SGX}
    \label{2epc}
    \vspace{-3ex}
\end{figure}

As we mentioned above, when SGX's secure memory is full, new memory allocation leads to page fault, which significantly brings down efficiency~\cite{arnautov2016scone}~\cite{costan2016intel}. Fig.~\ref{2epc} shows the time cost with respect to the increase in data size with or without Enclave. In both figures, red curve refers to the setting that the task is performed in SGX, where the vertical line refers to the standard size of Enclave under SGX, \ie 128\textsc{mb}. No matter whether it's a binary search or a quick sort, when the amount of data exceeds 128MB, or even less than 128MB, the cost of SGX increases significantly, while the execution outside Enclave (\ie blue curve) is unaffected. 

Therefore, combined with theoretical study and experimental results, it is more efficient to rely on TEE mode when no paging occurs, but when the secure memory is full and paging occurs, the efficiency can be reduced by 2-4 times. This makes it more expensive to continue with TEE-enabled mode.

\subsubsection{Self-adaptive mode switch}\label{sssecselfadap}
As analyzed above, limited space of Enclave will inevitably introduce too much paging task in high concurrency transactional scenarios. An ideal solution towards this problem is to obtain the state of the current SGX in real time and dynamically determine whether the current ciphertext predicates shall be performed in software mode or TEE-enabled one. That is, the system has to obtain the remaining capacity of the Enclave memory in real time to support the operation of the decision-making mechanism. Unfortunately, the official SDK of SGX does not provide such an interface to return the remaining capacity of enclave page caches(EPC) in real time. 

To address that, we present a micro-benchmark to estimate the residual capacity of EPC in real time. 
Intuitively, inspired by our study in Fig.~\ref{2epc}, if the remaining capacity of Enclave memory is insufficient, page replacement will occur, which will affect the execution efficiency significantly, then the running time of a UDF calculation task executed in SGX can indirectly reflect the status of the current EPC remaining capacity, that is, if the execution time of a predefined task in SGX is significantly higher than its expectation, then page replacement occurs during execution. In this way, the current remaining capacity of EPC can be dynamically inferred based on historical data.
In the sequel, we refer to this strategy as Enclave microbenchmark. Specifically, in a new thread, a timer is started with the system startup, and a task is triggered at a certain time interval. This task is executed in Enclave, and the execution time of this task is recorded. Because it is a fixed task, the running time is also relatively fixed.

For the specific type of benchmark test, users can fill in different tasks according to the performance of SGX platform, so as to obtain better estimation performance. One of the main demands of the benchmark task is that the access of the task test itself to the data should be random enough to resist against the caching effect of the page replacement algorithm, so that the task can cause enough page missing exceptions. In that way, differentiation in the running time can be easily observed. Generally, if the data access of the task is not random enough, there will be less page loss interruption, so it can not accurately reflect whether page replacement is happening at present. After exhaustive empirical study, which shall be shown in \autoref{ssec:expmicroben}, in Enc$^2$DB we adopt the binary search after quick sorting in Enclave memory as the benchmark task.

\subsubsection{cost estimation model}
The main task of the design of self-adaptive switch strategy is to dynamically provide the most appropriate calculation path for each UDF calculation according to the current system state. The execution mode for each particular predicate over ciphertext can be viewed as different physical operators (\ie software mode or TEE-enable mode, each corresponds to a UDF) in the query execution plan (QEP). We shall further deploy a cost model to enable the dynamic switch between physical operators, \ie UDFs. 
For ease of discussion, we refer to software and TEE-enabled mode UDF cost as $C_{soft}$ and $C_ {TEE}$, respectively. 
\begin{equation}
    C_{soft}=C_{calc}+C_{decide}
    \label{eq:pure_soft_cost}
\end{equation}
$C_{soft}$ is shown in equation \ref{eq:pure_soft_cost}, where $C_{Calc}$ is the calculation cost in UDF execution, $C_{decision}$ is the cost of the decision itself. Because the subsequent calculation involves the collection of the calculation time and the feedback to the decision-making, and the software and hardware belong to two different concurrent conflict domains, so $C_{decide}$ mainly depends on the number of UDFs of the same kind calculated on the same path at the current time. For different UDFs, $C_{Calc}$ is different. For instance, in the symmetric cryptographic homomorphic algorithm, the efficiency of AHE is always higher than that of MHE. An estimate of $C_{calc}$ can be given based on the computational flow of different UDFs and their complexity. In addition, compiled assembly instructions or runtime CPU clock cycles based on UDF code, can be used as a reference to assist in estimating $C_{calc}$.

And for $C_{TEE}$, the total cost can be found in equation \ref{eq:tee_cost}.
\begin{equation}
C_{TEE}=C_{fixed}+C_{calc}+C_{runtime}+C_{decide}
\label{eq:tee_cost}
\end{equation}

$C_{fixed}$ is the startup cost of the trusted execution environment. For instance, in SGX, it is mainly reflected as the $ECALL/OCALL$ invocation overhead. $C_{calc}$ is the computational cost in the UDF execution. Since the main logic of the execution in trusted hardware is to decrypt the AES, perform the computation on the plaintext, and then encrypt the result according to the encryption format required by the UDF, the dynamic part of the computational cost here depends on the parameters of the UDF, \ie the specified encryption scheme requirements. In decision making, it is necessary to estimate this part of the computational cost dynamically for different parameters. $C_{runtime}$ is the additional load overhead of the trusted hardware at runtime, \eg in SGX it is mainly expressed as the additional computational cost during page replacement. Its value varies continuously with the severity of page replacement. The main measurement method is to dynamically estimate $C_{runtime}$ from the microbenchmark test introduced above.

\section{FURTHER OPTIMIZATION}\label{sec4}
\subsection{cipher index}\label{ssec41}
B(+)-tree index is a fundamental tool to accelerate queries with either range or equivalent predicates.
In encrypted database, the client encrypts the numeric data and transmits the ciphertext as a string, which is typically stored in the ``text'' or ``blob'' type on the server side. Due to the different representation forms of plaintext and ciphertext, the classical comparison logic of plaintext database is no longer applicable over ciphertext. Correspondingly, B(+)-tree cannot directly support the query over ciphertext, \ie ORE in our software-only mode.

In this part, we present an index scheme supporting equivalent query, range query and related aggregate functions over ORE ciphertext that is compatible with both PostgreSQL and openGauss. 
Meanwhile, we ensure the transparency of user, who only need to enter \begin{lstlisting}[language=SQL,keywordstyle=\color{blue}
]
CREATE INDEX idx_name ON tb_name(col_name);
\end{lstlisting} to create server-side ciphertext indexes in the same way with plaintext databases. 

Intuitively, one possible solution to this index is to replace the comparison operator within B(+)-tree using ORE-based UDF that can return comparison results over ORE ciphertext. However, cost model and query optimizer is unaware of the cost of this modified B(+)-tree, resulting in the fact that the query plan may not correctly employ this index, \ie wrongly use \textsf{Seq SCAN} instead of \textsf{Index SCAN}. To address that, our model relies on UDT (user-defined type) and UDO (user-defined operation) instead of purely UDF over ORE ciphertext.
\subsubsection{user-defined data type}
Firstly, we define ``ore\_en'' type as a variable-length data type ``varlena'' provided by PostgreSQL (\resp openGauss). 
In addition, since the length of the ore ciphertext is several times or even dozens of times the length of the plaintext, it is necessary to mark the type as TOAST, so that the ciphertext can be compressed and stored off the line.
\subsubsection{user-defined operators}
Range query consists of the following operators, ``$>, \ge, =, <, \le$'', each of which is a binary operation a pair of ORE ciphertext (\eg A,B) defined as type ``ore\_en'', returning either 1 or -1 as follows:
\begin{enumerate}
    \item \textsf{ore\_en\_abs\_lt(A, B)}: returns 1 if 
    $\mathsf{Dec(A)}<\mathsf{Dec(B)}$.
    \item \textsf{ore\_en\_abs\_le(A, B)}: returns 1 if 
    $\mathsf{Dec(A)}\le\mathsf{Dec(B)}$.
    \item \textsf{ore\_en\_abs\_gt(A, B)}: returns 1 if 
    $\mathsf{Dec(A)}>\mathsf{Dec(B)}$.
    \item \textsf{ore\_en\_abs\_ge(A, B)}: returns 1 if 
    $\mathsf{Dec(A)}\ge\mathsf{Dec(B)}$.
    \item \textsf{ore\_en\_abs\_eq(A, B)}: returns 1 if 
    $\mathsf{Dec(A)}=\mathsf{Dec(B)}$..
\end{enumerate}
As an alternative solution, the equivalent predicate can be also performed through the DET ciphertext column, \ie without our cipher index.

According to the comparison functions constructed above, five different operators can be created correspondingly. In both PGSQL and openGauss, it can be done via ``\textsf{CREATE OPERATOR}'' statement over the new UDT, namely ``ore\_en''.

Essentially, given that the ``ore\_en'' data type is associated with a series of comparison operators, \ie ``$>,<,\le,\ge,=$'', a B(+)-tree index can be easily defined accordingly. 

\subsubsection{new operator class}
The operator class can inform the B-tree index of which data type to operate on, providing a set of operations supported by a certain data type and implementing B-tree access for new types. In PGSQL and openGauss, different operator classes and index methods can be built for different types. Every index defines its own support function for comparison or operation logic. 

For UDT ``ore\_en'', we define a B-tree operator class named ``ore\_en\_abs\_ops''. In the B-tree index, the five comparison operations are referred to using IDs as ``$1,\ldots,5$'', respectively. In the operator class of ``ore\_en'', by specifying the UDO as the ciphertext comparison operator above, an ciphertext B-tree can be built upon the UDT ``ore\_en'' and the associated UDO class, which consists of the UDOs defined above. 

In addition, the above ciphertext index has also brings in a new advantage to our software-only mode implementation. Before introducing the index, performing sorting over ORE columns may inevitably result in wrong output, as the \textsf{SORT} operator has to rely on the default ASCII code of the ciphertext, \ie ``order by'' statement in SQL fails to work. By implementing the cipher index, \textsf{SORT} operator over the UDT ``ore\_en'' is now conducted based on the UDO, ``$\le$ over ore\_en'', \ie \textsf{ore\_en\_abs\_le(A, B)}.
\subsubsection{Ciphertext range query}
By introducing the UDT, UDOs above, a range query over ORE columns will not rely on the UDFs introduced in \autoref{udfs} ever, \eg ``...WHERE udf\_ore\_gt(...);''. Instead, the range predicate turn into the form of ``$<,\le,>,\ge,=$'', which is inline with plaintext SQL. The advantage of this is two-folds, firstly, it ensures the range query to employ the cipher index to take effect (the default solution shown in Fig.~\ref{udfs} can not correctly use \textsf{Index SCAN}); secondly, the appearance and usage of the range predicates are consistent with the plaintext ones, ensuring the transparency over users.

Besides, as the maximum/minimum aggregation depend on comparison and sorting over the ORE column, which are now stored using UDT ``ore\_en'', the aggregate function on ``ore\_en'' data type are also introduced and defined as ``ore\_max(col\_ore)/ore\_min(col\_ore)''. 

\subsection{within-SGX caching}
\begin{figure}[t]
    \centering
    \includegraphics[width=1\linewidth]{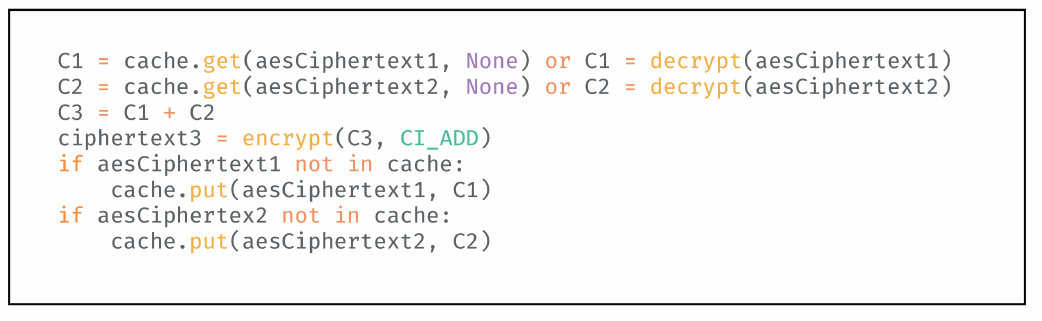}
    \vspace{-4ex}\caption{Pseudocode for within-SGX cache}
    \label{fig.4.7}\vspace{-3ex}
\end{figure}

In the design of cryptographic database system, cipher calculation is costly, so it is a waste of resource especially when repeated query (\resp sub-query) happens. If we can reduce the duplicate cipher calculation, it will not only improve the efficiency, but also save much secure memory space. As AES ciphertext are repeatedly decrypted to perform predicate computation within SGX, it is possible to cache the cipher-to-plain text mapping of AES within SGX, such that the follow-up decryption task towards the same AES ciphertext can be accelerated significantly. Fig.~\ref{fig.4.7} shows the pseudocode how we implement ciphert-to-plain caching in SGX.
LRU or LFU can be potential choice for implementing the cache algorithm. As cached data exists in limited secure memory, the cache algorithm also needs to save memory as much as possible, so LRU is a more appropriate choice. Experiments show that ciphertext caching has a 15\%-20\% improvement over TPC-C benchmark. For the aspect of cache capacity, in Enc$^2$DB we set it configurable, such that it can be adjusted according to different usage scenarios and properties of different SGX platforms.

\subsection{SGX task pool for batched ecall}
In an Enclave program, access to the enclave program from the untrusted side is done through a predefined bridge function called $\mathsf{ECALL}$. The verification of $\mathsf{ECALL}$ is also an computationally expensive process, so if one $\mathsf{ECALL}$ call enters the Enclave, whose computation task is too simple, the verification of $\mathsf{ECALL}$ accounts for most of cost, especially in face of a large number of concurrent short tasks. Hence, it is important to reduce the number of $\mathsf{ECALL}$ with respect to petty tasks. Notably, SGX has proposed \textit{Switchless Call}~\cite{SGE-SDK} technique to reduce this type of overhead by deploying worker threads inside the Enclave to asynchronously acquire tasks for execution. However, according to \textit{Switchless Call}, work threads can only be specified statically and the size of the task pool is fixed, which makes it impossible for cryptographic database systems to achieve more fine-grained adjustment and optimization according to the characteristics of ciphertext (UDF) computation tasks.

To address that, we deploy a $\mathsf{ECALL}$ task pool at the untrusted end. Whenever a thread calls $\mathsf{ECALL}$, a task object, containing its ID and parameters, is generated and stored in the task pool. The work thread at the trusted end takes the $\mathsf{ECALL}$ task from the task pool and executes it. After a certain number of tasks in the pool or a certain time window, all tasks are transferred to the processing pool, at which time the work thread in the safe zone will iterate through the tasks from the processing pool and execute them accordingly. After all tasks in the processing pool have been executed, new tasks from the pool can be accepted. The reason for dividing the pool into two sub ones is that threads in the safe zone cannot share the same lock mechanism with threads in the non-safe zone, because the lock object for non-safe threads is provided by the standard library, while the lock object for safe zone threads is provided by the SGX development library.
This pattern effectively prevents the $\mathsf{EENTER}$/$\mathsf{EEXIT}$ instructions from being called, thus reducing additional overhead. Notably, when the task pool is full or all work threads are busy, the $\mathsf{ECALL}$ call degenerates to a normal form of call.

In addition, a more important point is that because \textit{Switchless Call} is a multi-threaded model, and a multi-process model database system like PostgreSQL cannot effectively use \textit{Switchless Call} technology and must implement a inter-process concurrency control instead, this is why the proposed task pool is essential.

\section{Evaluation}\label{sec5}
We verify the performance of our system in both software-only mode and TEE-enabled mode on TPC-C benchmark implemented using sysbench\cite{kopytov2017sysbench} with test script Sysbench-TPCC\cite{tpccsysbench}. The study is conducted on an Intel SGX-enabled machine with an EPC size of 128MB, equipped with an 8-core, 16-thread Intel Xeon E-2288G CPU with 512KB, 2MB and 16MB of L1,L2,L3 cache, respectively. We compare with a pair of baselines, including the original (plaintext) implementation of the same database, \ie PostgreSQL/openGauss\footnote{We implement Enc$^2$DB on both PostgreSQL and openGauss, as the results on both system are consistent we select to showcase only that of  PostgreSQL due to space limit and popularity among the audience.}, and the state-of-the-art software-based encrypted database, namely Symmetrial~\cite{savvides2020efficient}. Since the TPC-C test cannot manually control the read/write ratio, we also perform experiments over synthetic dataset (1 million records) by allowing configurable read/write ratio. The specific list of hardware/software experimental information is shown in Table \ref{exp}.

\begin{table}[t]
    \centering
    \caption{Experimental configuration list}
    \label{exp}
    \begin{tabular}[t]{ll}
    \toprule
    Item & configuration\\
    \midrule
        CPU  & Intel(R) Xeon(R) E-2288G CPU @ 3.70GHz \\ 
        Memory & 64GB \\
        Storage & Samsung SSD 840 Pro @ 512GB \\
        OS & Ubuntu 18.04 LTS \\
        PostgreSQL & 14.1 \\
        Intel SGX SDK & 2.15.101 \\
        Intel SGX Driver & 1.41 \\
    \bottomrule
    \end{tabular}
  \end{table}

\subsection{Overall throughput}
\begin{figure}
  \centering
  \subfigure[QPS]{\label{5overall:a} \includegraphics[width=.48\linewidth]{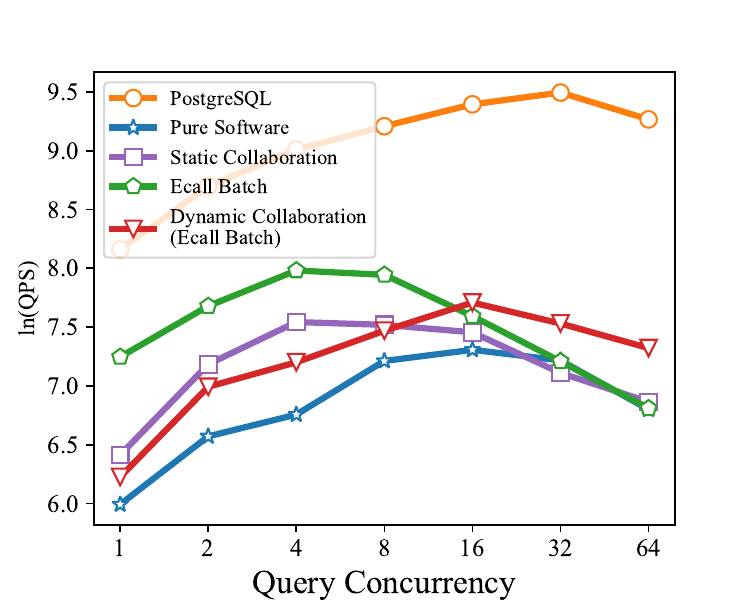}}
  \subfigure[TPS]{\label{5overall:b} \includegraphics[width=.48\linewidth]{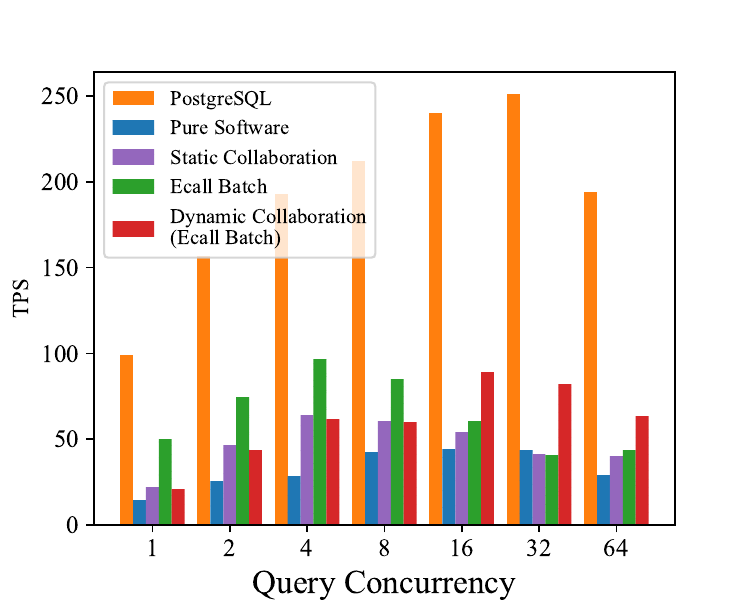}}
  \caption{Overall results on TPC-C}
  \label{5overall}
  \vspace{-4ex}
\end{figure}

The overall performance in the aspects of both latency and TPS (\#Transactions per second) by varying the scale of concurrency is shown in Fig.~\ref{5overall}, covering native (plaintext) system, software-only mode, static (mode switch) TEE-enabled mode w/wo task pool, and dynamic (mode switch) TEE-enabled mode. Since the system CPU contains 16 cores, all solutions show a decrease in QPS when the concurrency number exceeds that of CPU cores. The software-only mode constantly exhibits $<10\times$ performance loss compared to the plaintext database. In the static TEE-enabled mode, since they only do a simple replacement, a more severe page replacement occurs in the scenario with higher concurrency, and their performance is not even as good as the software-only mode at high concurrency. In the scenario with low concurrent requests, the performance is better, and the loss is $<5\times$ compared to the plaintext database. In low concurrency scenarios, task pooling performs the best among all solutions due to the elimination of the computational overhead caused by $\mathsf{ECALL}$s. The task pool mode are more advance for multi-threading, but performs not so good in multi-process tests. However, since PostgreSQL's concurrent implementation is multi-process mode, it does not perform as well in high concurrency scenarios as the static TEE-enabled mode without pool. The dynamic TEE-enabled mode does not perform as well as the other two, \ie static TEE-enabled mode w/wo task pool, in the lower concurrency scenarios, because the low incidence of page replacement. Instead, it performs best in the high concurrency scenarios where page replacement frequently happens.

\begin{figure}
  \centering
  \includegraphics[width=1\linewidth]{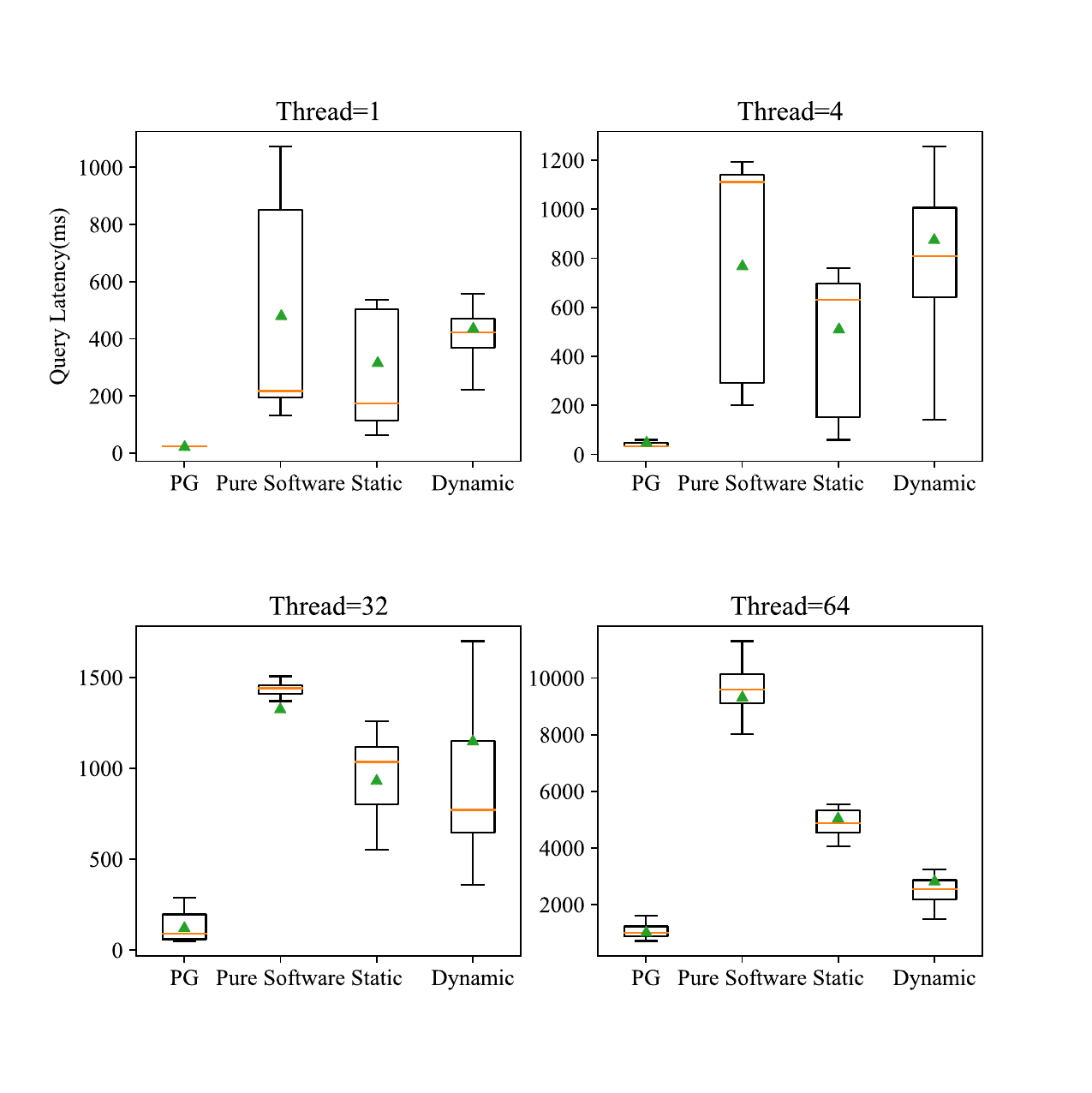}
  \vspace{-10ex}
  \caption{TPC-C Query Latency Variation}
  \label{5boxplot2.pdf}\vspace{-1ex}
\end{figure}

The TPC-C test provides feedback on the latency of query execution for each statement, and the results are shown in Fig.~\ref{5boxplot2.pdf}. As SGX task pool shows advance performance in all TEE-enabled mode, in the rest experiments it is by default turned on in TEE-enabled mode solutions. Fig.~\ref{5boxplot2.pdf} collects the average query latency every 10 seconds and present the data collected under four different concurrency settings. The upper and lower edges of the box in the figure indicate the upper and lower quartiles in a batch of data respectively, and the upper and lower boundary are the observed extreme values respectively. The red line refers to the median, the triangle is the mean.

The mean and median of the static TEE-enabled mode are the lowest when concurrency is between 1 and 4, while the dynamic TEE-enabled mode is better in the high concurrency scenarios (32 to 64). Although dynamic TEE-enabled mode does not perform as well as the static one at low concurrency, the query latency is lower than both software-only mode and static TEE-enabled mode at high concurrency, indicating that dynamic scheduling execution of TEE plays a more important role. Numerically, as the number of concurrency increases, all solutions show a trend of increase with concurrency, and this phenomenon is more significant in TEE-enabled mode. It is possibly caused by the fact that a more severe page substitution occurred in the experiments, which led to a sharp increase in query latency. In addition, the average in the experiments is generally higher than the median, and this phenomenon is especially obvious in the dynamic TEE-enabled mode, indicating that some queries are more time-consuming, mainly because the TEE-enabled mode solutions need to do additional inter-process synchronization work during initialization, and the computational overhead of initialization affects the latency of the corresponding queries.

\subsection{Self-adaptive mode switch test}\label{ssec:expmicroben}
\begin{table}
  \centering
  \renewcommand{\arraystretch}{1.5}
  \caption{Impact of probing task type and data size on QPS}
  \label{detector.tb}
  \begin{tabular}[t]{|c|c|c|r|}
  \toprule 
  \multicolumn{1}{|l}{\diagbox[innerwidth=2cm]{Data Size}{Task Type}} &\multicolumn{1}{|c|}{Binary Search}&\multicolumn{1}{c|}{Quick Sort} & \multicolumn{1}{r|}{Mixed Task}\\
  \hline
  1024&1564.41&1690.49&1666.71\\
  \cline{1-4}
  8192 & 1753.92 & 1694.29 & \pmb{1830.58*}\\
  \cline{1-4}
  12288 & 1666.71 & 1660.97 & 1736.01\\
  \cline{1-4}
  16384 & 1661.77 & 1688.45 & 1680.66\\
  \bottomrule 
  \end{tabular}
\end{table}

\begin{figure}
  \centering
  \subfigure[thread=4]{\label{5detect1} \includegraphics[width=.48\linewidth]{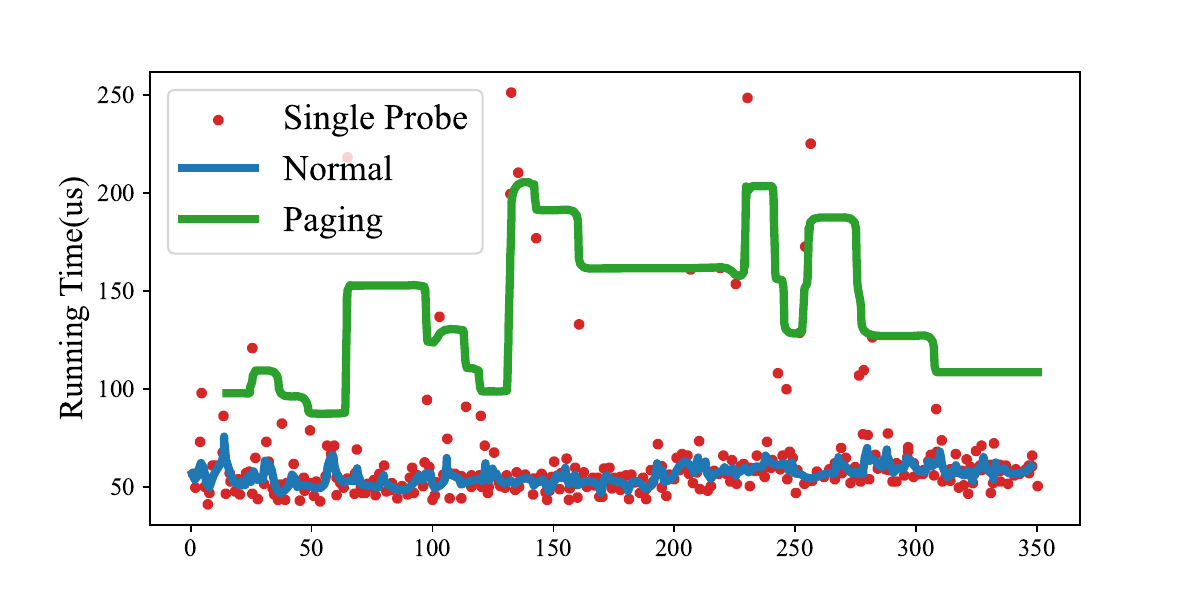}}
  \subfigure[thread=64]{\label{5detect2} \includegraphics[width=.48\linewidth]{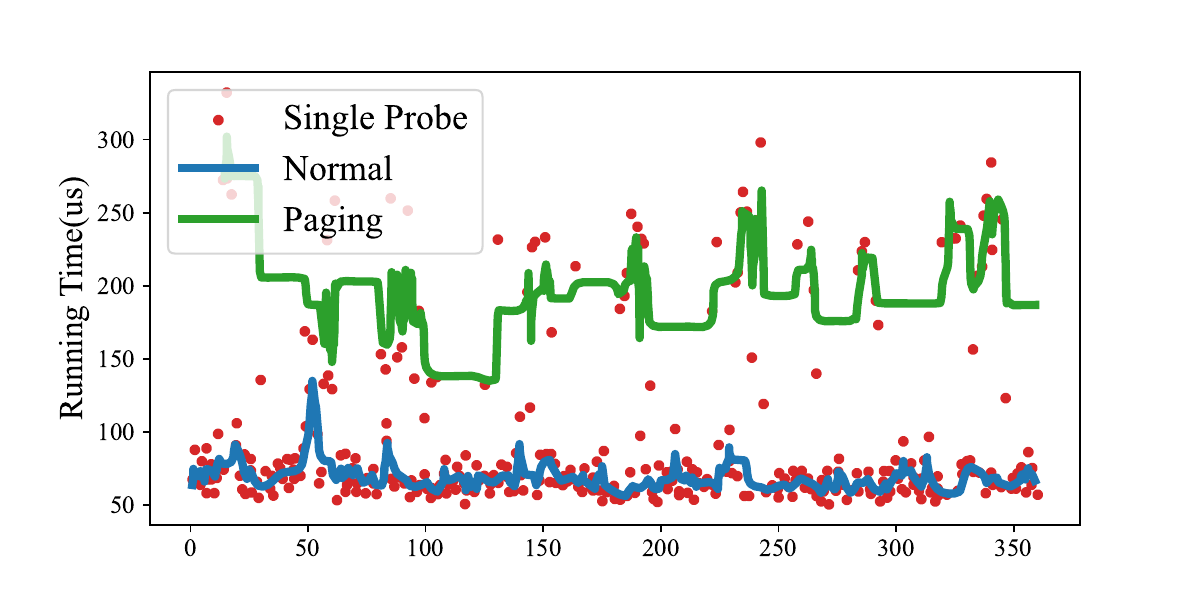}}
  \caption{Running Time of Probing Task during TPC-C}
  \label{5detect}\vspace{-3ex}
\end{figure}

In dynamic mode switch scheme of TEE-enabled mode show in \autoref{sssecselfadap}, we present a micro-benchmarking to estimate the state of the current EPC operation, so that to switch between ciphertext predicate solution self-adaptively. Fig.~\ref{5detect} shows the time cost for micro-benchmark task during the TPC-C test, where the scatter refers to the time of each micro-benchmark task execution, and the dash indicates the average running time of the current mode with (replacement state) or without page replacement (normal state), respectively. Obviously, the execution time of the micro-benchmark task in the replacement state is in general twice as long as that in the normal state. Fig.~\ref{5detect1} and Fig.~\ref{5detect2} are the results under different concurrency settings, respectively. By comparing between both figures, it can be seen that in the high concurrency scenario, the probing tasks take more time in the replacement state than in the low concurrency scenario. The estimation results are in line with the ground-truth state inside the Enclave, hence it can inform us the real-time state of secure memory for self-adaptive mode switch.

In addition, we also conduct experiments to explore the performance by varying the types of micro-benchmark probing and task workload volumes. As shown in \autoref{detector.tb}, three different task types are considered, namely, binary search, quick sort and mixed tasks (containing both), over four task workload volumes. The results show that for the current test, all task types show the best performance with 8192 data size, and mixed tasks perform best among all types.

\begin{figure}
  \centering
  \includegraphics[width=1\linewidth]{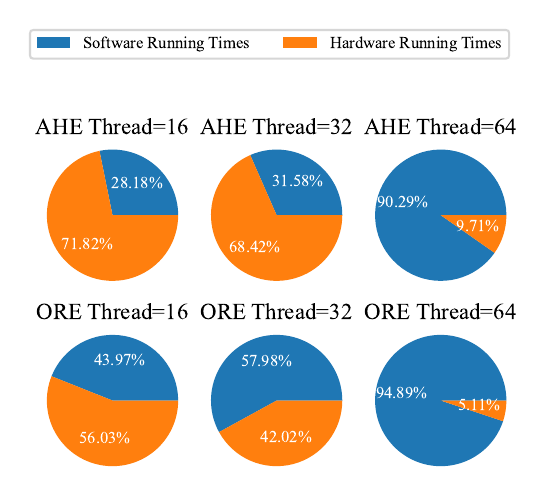}
  \vspace{-6ex}
  \caption{Proportion of time-consumption for different phases under TEE-enabled mode with self-adaptive switch strategy}
  \label{5co_pie}
  \vspace{-3ex}
\end{figure}

Fig.~\ref{5co_pie} shows the number of times the homomorphic addition and ORE are executed under different concurrency settings. Obviously, with the increase of concurrency, the SGX memory space is gradually occupied, at this time, under the regulation of the self-adaptive mode switch, the UDF calculation using Enclave becomes less, and more UDFs are calculated outside Enclave, via software-only mode. The main reason is that in high concurrency scenario, too many UDFs enter the feasible hardware at the same moment, resulting in more data to be processed simultaneously within it, which exceeds the threshold of page replacement, and the micro-benchmark probing task execution time will increase. With the help of the micro-benchmark, Enc$^2$DB dynamically schedule the execution mode of UDFs, so as to control the use of EPC space to reduce the extra overhead of page replacement.

\begin{figure}
  \centering
  \includegraphics[width=1\linewidth]{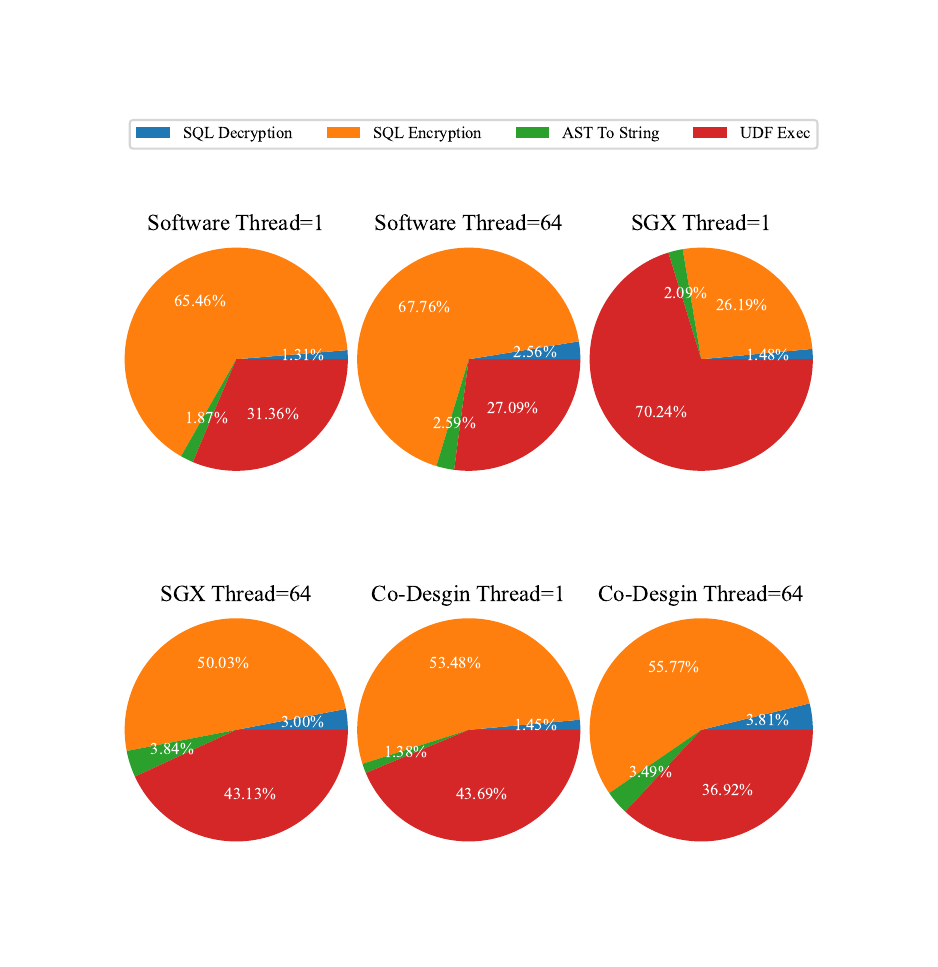}
  \vspace{-8ex}
  \caption{Time-consumption between components in Enc$^2$DB}
  \label{5part_pie.eps}
  \vspace{-1ex}
\end{figure}

The percentage of time consumed by different phases within Enc$^2$DB under TPC-C test is shown in Fig.~\ref{5part_pie.eps}. In software-only mode, the most time-consuming module is SQL Encryption, which is all cryptographic operations. Encryption generally occurs on the client side and triggers whenever a query is stated, especially for insert statement, which requires encryption over a set of entries. Due to the high proportion of insert operations in TPC-C tests, Encryption dominates in the pie chart. If there are more pure read request, its proportion will be reduced. The second most time-consuming part is the UDF predicate execution, which reflects the proportion of ciphertext calculations involved in the query. It is worth noting that in the implementation of static mode switch, when the concurrency number is 1, UDF evaluation takes more time than Encryption. The reason is that after replacing ORE with AES (under TEE-enabled mode), all plaintext columns now correlate to exactly one (AES) column (instead of 4 \ie AHE, MHE, ORE and AES), so the required encryption operation time will be reduced significantly. This is also one of the important reasons why the efficiency of TEE-enabled mode with static mode switch strategy is higher than that of software-only mode.

\begin{figure}
  \centering
  \includegraphics[width=1\linewidth]{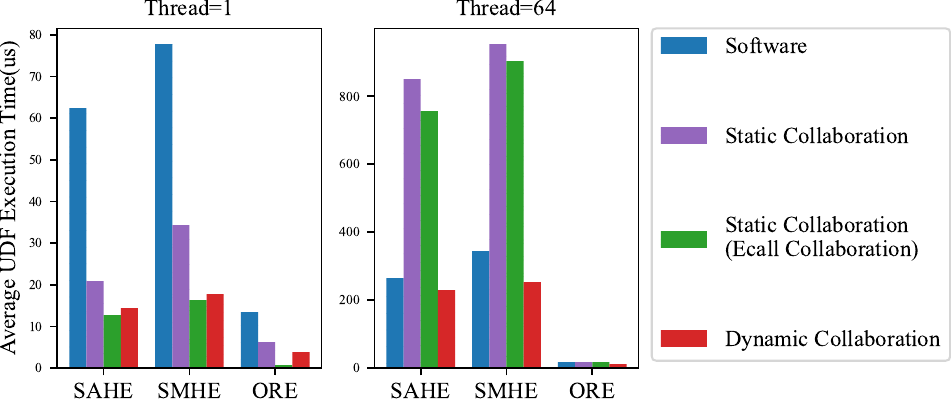}
  \vspace{-4ex}\caption{Time consumption for different predicates under different modes}
  \label{5udf.eps}
  \vspace{-2ex}
\end{figure}

In addition, we further tested UDF execution times for different ciphertext computation, including AHE for additive predicate, MHE for multiplication predicate and  ORE for comparison predicate over two extreme scenarios, \ie single thread and 64 threads, as shown in Fig.~\ref{5udf.eps}.

\begin{figure*}[t]
  \centering
  \subfigure[cache module choice]{\includegraphics[width=.48\columnwidth,height=2.7cm]{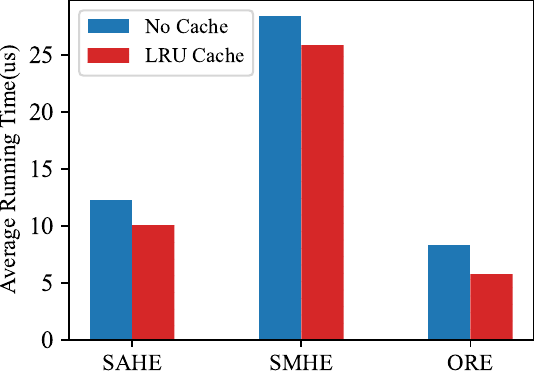}%
  \label{5cache}}
  \subfigure[cache hit]{\includegraphics[width=.48\columnwidth,height=3cm]{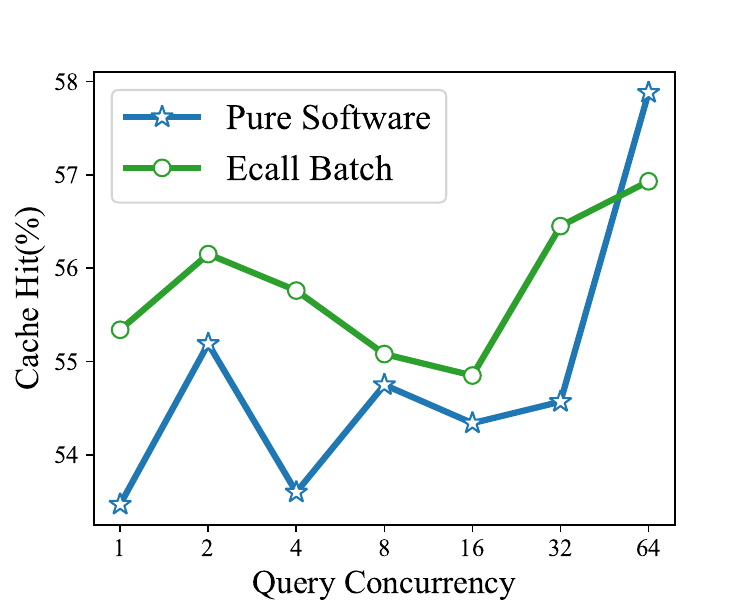}%
  \label{cache_hit1}}
  \subfigure[Cache replacement]{\includegraphics[width=.48\columnwidth,height=3cm]{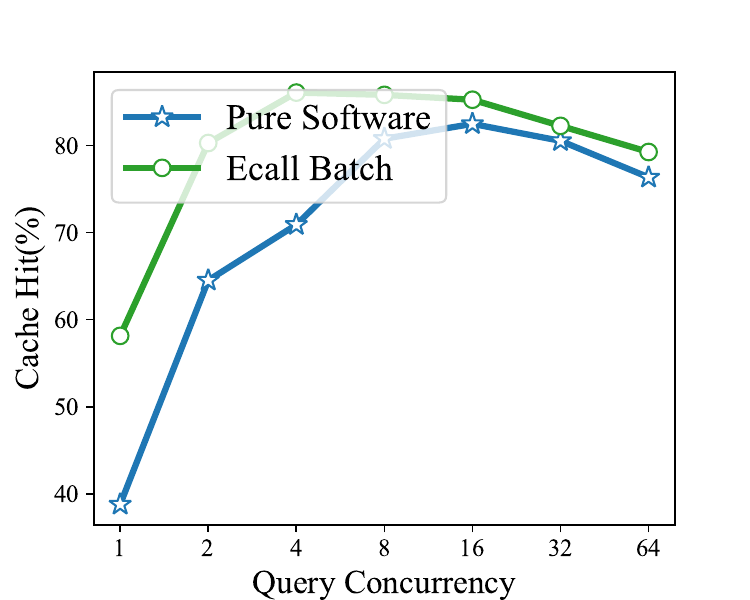}%
  \label{cache_hit2}}
  \subfigure[Storage space]{\includegraphics[width=.48\columnwidth,height=2.7cm]{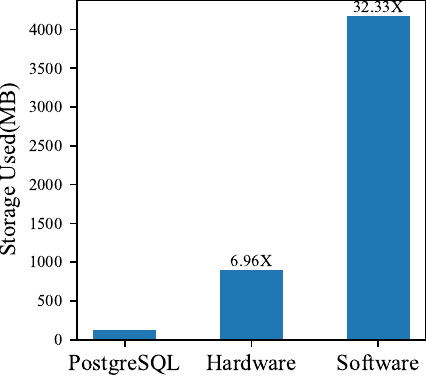}%
  \label{5storge}}
  \vspace{-1ex}\caption{(a) With-in SGX cache choice; (b-c) client-side cache effect; (d) storage expansion}
  \label{cache_hit}
  \vspace{0ex}
\end{figure*}

In the case of a single thread, \ie left of Fig.~\ref{5udf.eps}, no paging in SGX is triggered, static mode switch with task pool performs the best, mainly because the extra overhead of $\mathsf{ECALL}$ is eliminated in this mode. In comparison, the self-adaptive mode switch with task pool performs worse than the static one. This means that the execution path will dynamically select the TEE to realize at this time, but the micro-benchmark probing itself will incur time overhead.

In the case of high concurrency setting, where the client initiates 64 connection requests at the same time, the situation is reversed because page substitution occurs for static mode switch strategy, resulting in a significant discount in efficiency. At this time, regardless of whether the task pool mode is turned on or not, the computational overhead is about four times that of software-only mode. In this group of experiments, self-adaptive mode switch is the most efficient solution because TEE can be dispatched dynamically, which minimizes the cost of page substitution while making full use of the advantages of software computing. 

\subsection{Effect of within-SGX cache}\label{ssec:cache}
We have also conducted experiments to test the overall system performance improvement of the within-SGX cache module, which is deployed in secure memory to cache AES decrypted data. The experimental results are shown in Fig.~\ref{5cache}. From the experiment results, the cache module can provide 10\%-40\% efficiency improvement.


Besides the cache module deployed in the trusted hardware of the server, which eventually improves the efficiency significantly as shown in Fig.~\ref{5cache}. Inspired by that, we also try to deploy a cache module in the client's AES decryption component and explore its performance, details are shown in Fig.~\ref{cache_hit1} and Fig.~\ref{cache_hit2}.
Fig.~\ref{cache_hit1} shows the hit rate of the cache, \ie the probability of hitting the cache each time cache data is requested. Its performance is around 50\%, and the improvement is not obvious enough. Fig.~\ref{cache_hit2} shows the replacement rate of the cache, that is, the probability that the cache is full and the old data needs to be replaced each time the cache is missed. The higher the rate, the worse the cache effect. The replacement rate on the client side is also too high and the effect is not ideal. The main reason is that most of the decrypted data are expression results, the randomness of which is too large. Therefore, we select not to implement the AES decryption cache on the client side in Enc$^2$DB.
\begin{figure*}[t]
  \centering
  \subfigure[TPS in read/write balanced case]{\includegraphics[width=.48\columnwidth]{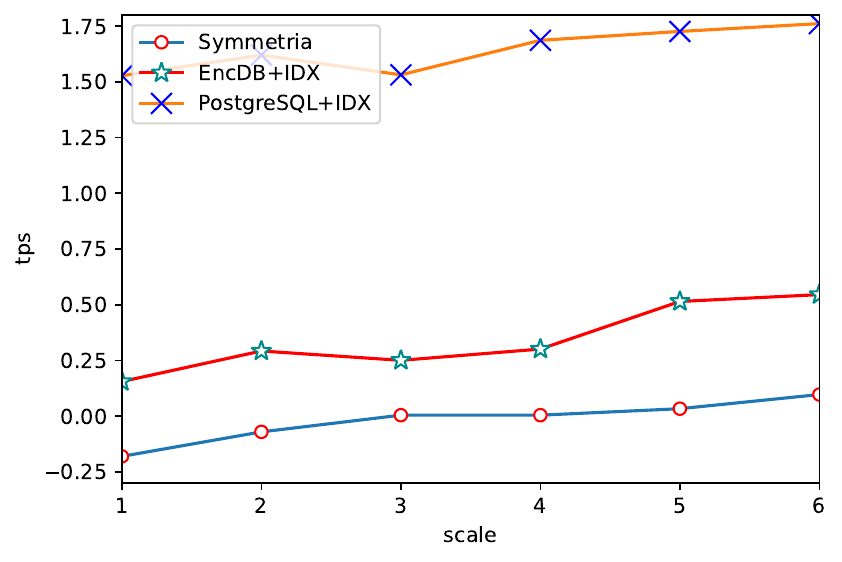}%
  \label{fig:TPC-C1}}
  \subfigure[TPS in read-only case]{\includegraphics[width=.48\columnwidth]{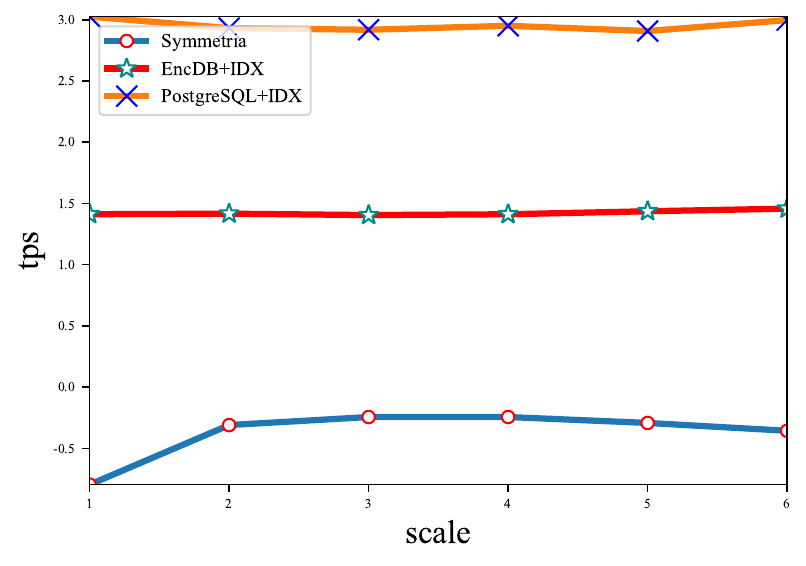}}%
  \label{fig:TPC-C4}
  \subfigure[TPS in multithreading read/write balanced case]{\includegraphics[width=.48\columnwidth]{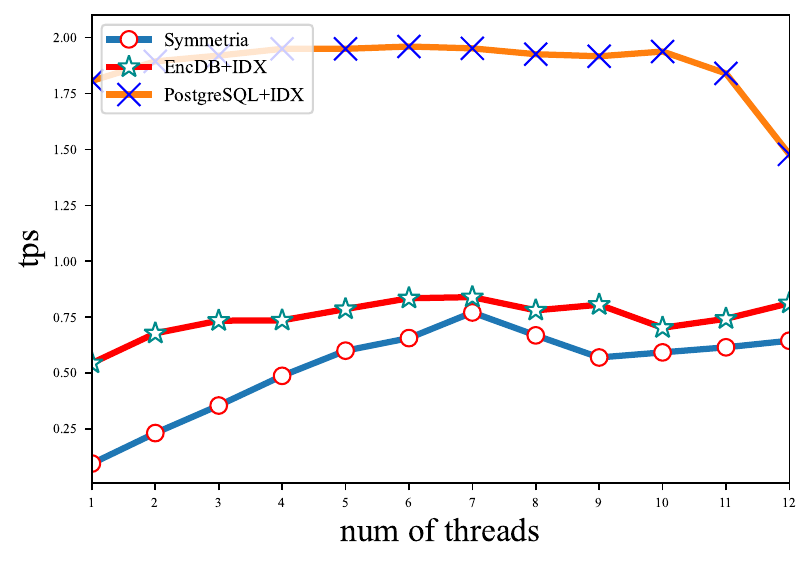}%
  \label{fig:TPC-C1_thread}}
  \subfigure[TPS in multithreading read-only case]{\includegraphics[width=.48\columnwidth]{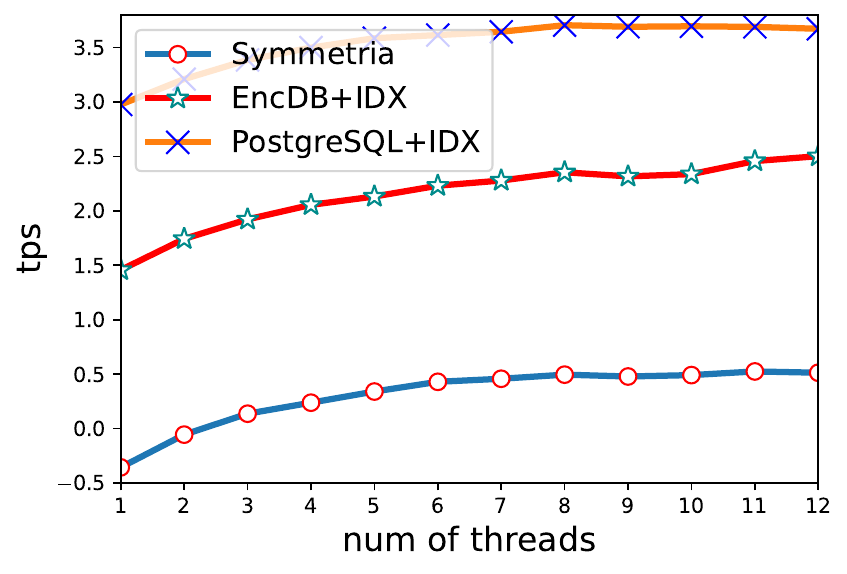}%
  \label{fig:TPC-C4_thread}}
  \caption{TPS in different scenarios}
  \label{fig:TPS}\vspace{-1ex}
\end{figure*}

\subsection{Storage expansion}\label{ssec:storageexp}
Beside, we also compare the storage space expansion of our solutions. The experiment is set to the TPC-C test script with parameters $table = 1$ and $scale = 1$. The specific experimental data are shown in Fig.~\ref{5storge}.

In software-only mode, the expansion of storage space is at least 30 times larger than that of the original database, while in the TEE-enabled mode, the expansion of storage space is about 7 times. The main reason is that in the TEE-enabled mode, the ORE encryption columns, whose ciphertext takes up too many bytes, are eliminated.

\subsection{Effects of cipher index in software-only mode}
For ease of discussion, we name software-only mode as EncDB and the mode with cipher index introduced in \autoref{ssec41} as EncDB+IDX. We test the performance of EncDB+IDX, Symmetria and PostgreSQL under different table capacities and thread numbers in read-only and read/write  workloads. In addition, we also record the QPS corresponding to the independent execution of different types of SQL statements (read/write), so as to evaluate the performance of different solution in the specific SQL execution.

\begin{figure}[t]
  \centering
  \includegraphics[width=.9\linewidth]{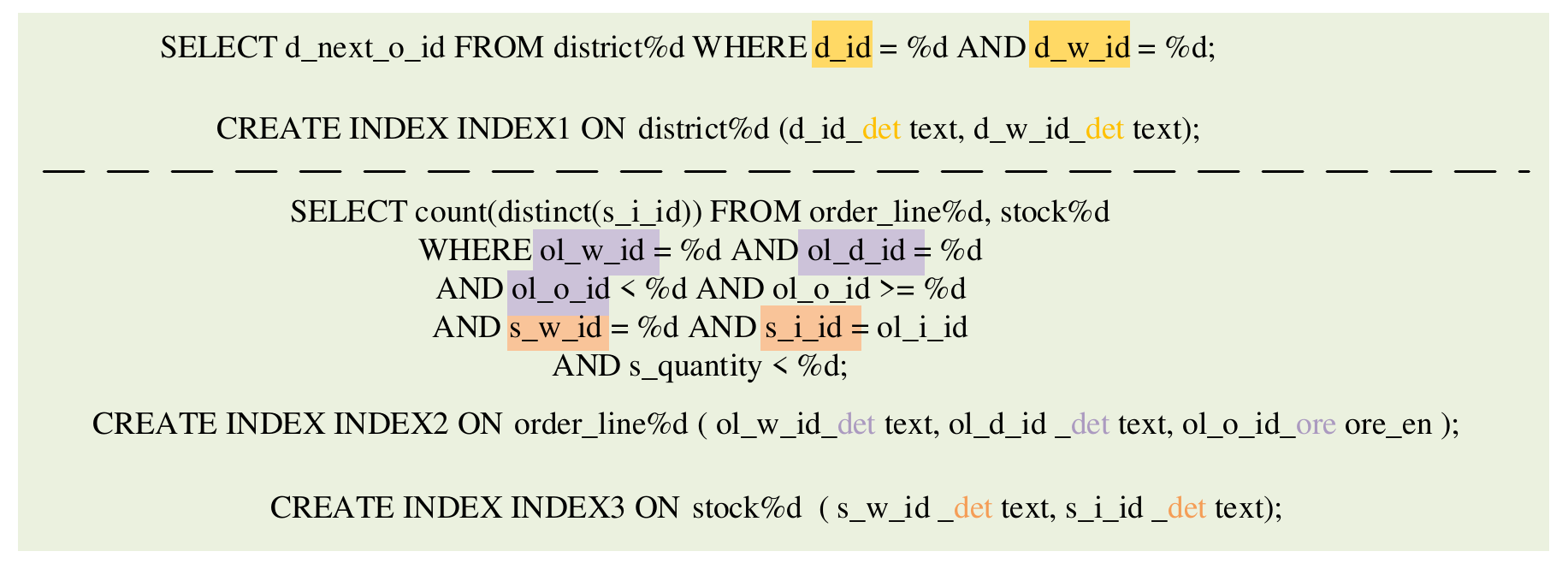}
  \vspace{-1ex}\caption{Read-only query statement}
  \label{index_location}\vspace{-1ex}
\end{figure}

The read-only test is carried out with respect to the ``stock level'' transaction, and the proportion of equivalent query and range query (an example can be seen in Fig.~\ref{index_location}) execution is 1:1. In the read/write balancing scenario, all transactions supported by TPC-C are executed, including read/write statements.

\begin{figure*}[t]
  \centering
  \subfigure[QPS in read/write balanced case]{\includegraphics[width=.48\columnwidth]{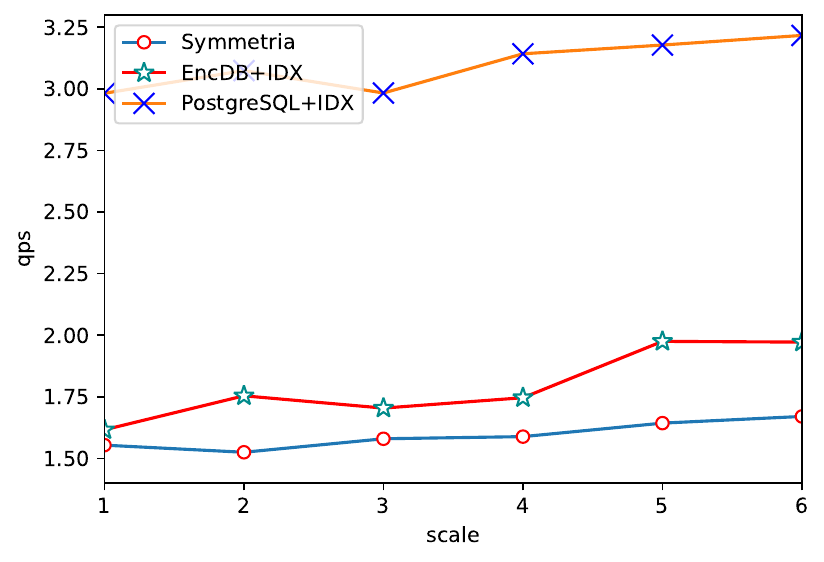}%
  \label{fig:TPC-C2}}
  \subfigure[QPS in read-only case]{\includegraphics[width=.48\columnwidth]{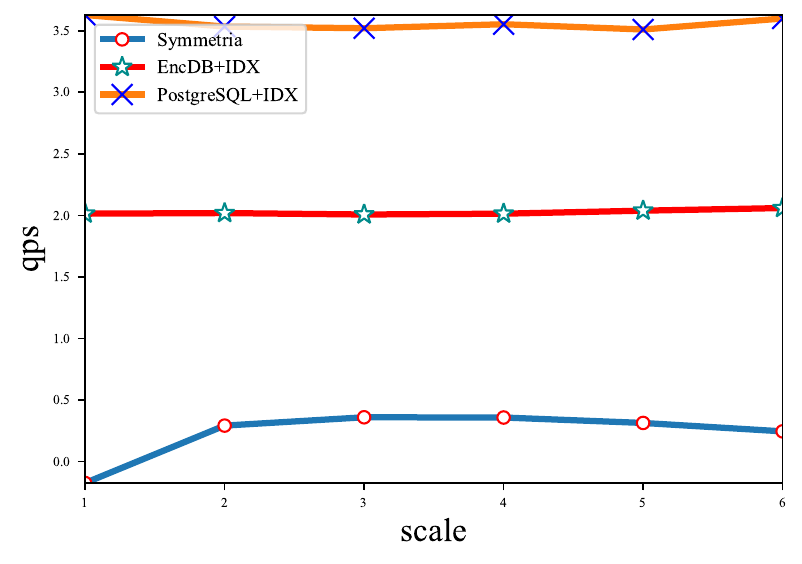}%
  \label{fig:TPC-C3}}
  \subfigure[QPS in multithreading read/write balanced case]{\includegraphics[width=.48\columnwidth]{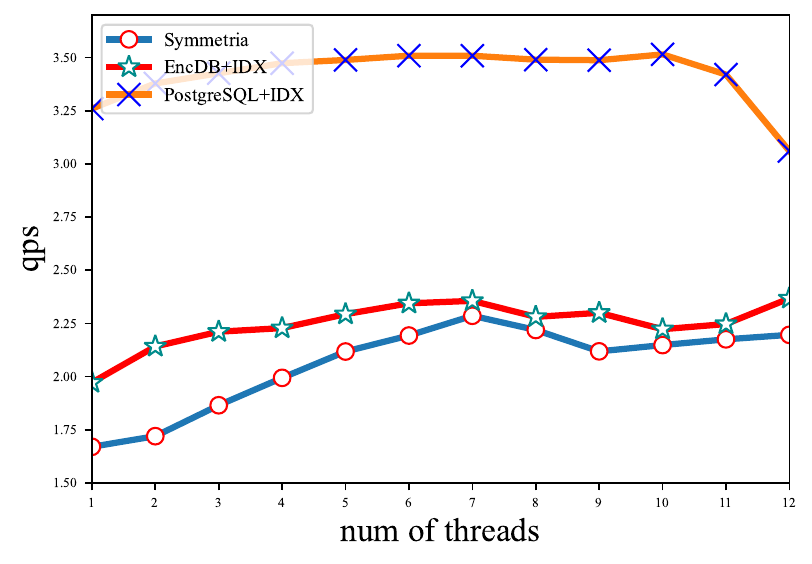}}%
  \label{fig:TPC-C2_thread}
  \subfigure[QPS in multithreading read-only case]{\includegraphics[width=.48\columnwidth]{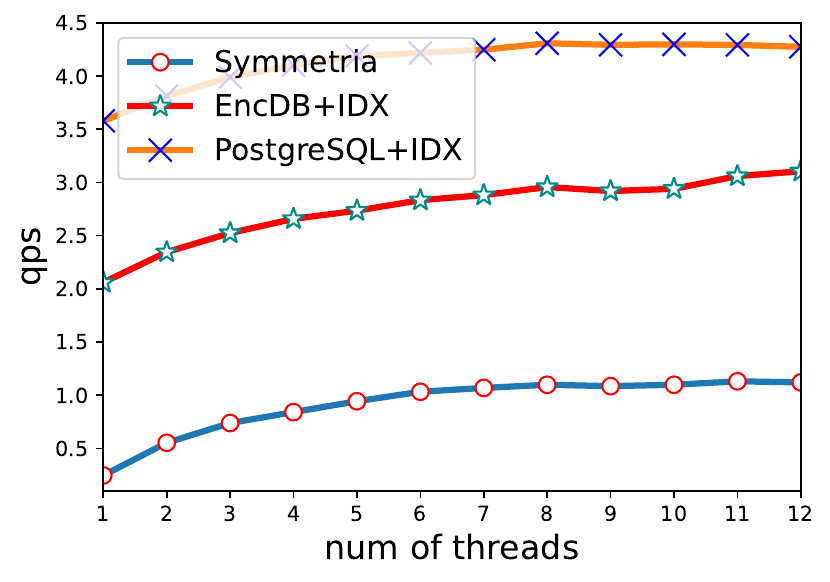}}%
  \label{fig:TPC-C3_thread}
  \vspace{0ex}\caption{QPS in different scenarios}
  \label{fig:latency}\vspace{0ex}
\end{figure*}
\begin{figure*}[t]
  \centering
  \subfigure[latency in read/write balanced case]{\includegraphics[width=.48\columnwidth]{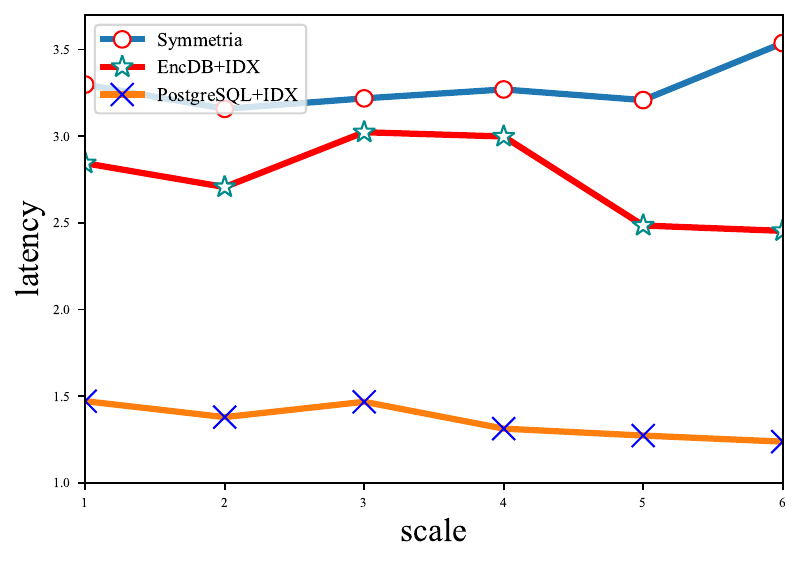}}%
  \label{fig:TPC-C5}
  \subfigure[latency in read-only case]{\includegraphics[width=.48\columnwidth]{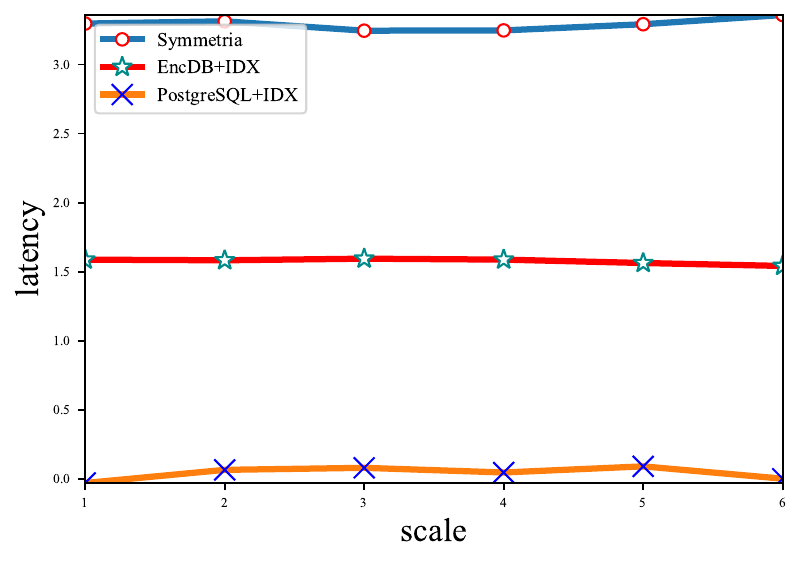}}%
  \label{fig:TPC-C6}
    \subfigure[latency in multithreading read/write balanced case]{\includegraphics[width=.48\columnwidth]{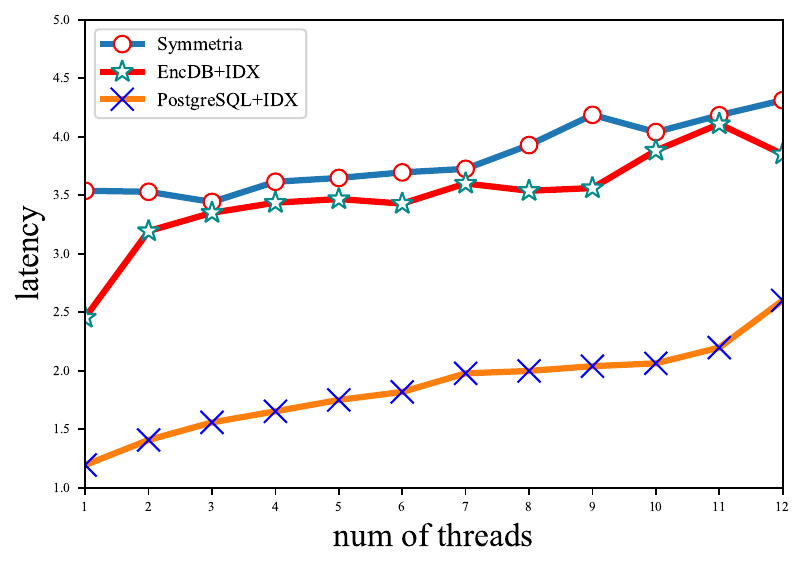}}%
  \label{fig:TPC-C5_thread}
  \subfigure[latency in multithreading read-only case]{\includegraphics[width=.48\columnwidth]{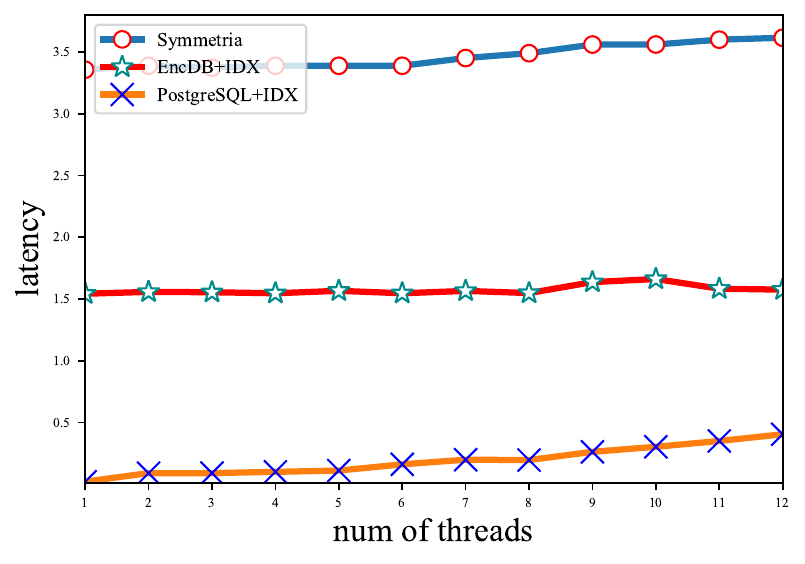}}%
  \label{fig:TPC-C6_thread}
  \vspace{0ex}\caption{latency in different scenarios}
  \label{fig:QPS_thread}
  \vspace{0ex}
\end{figure*}
A joint index for the equal predicate columns corresponding to ``d\_id" and ``d\_w\_id" is established on the ``district'' table. The ``order\_line'' table constructs a joint index for ``the DET column of ol\_w\_id, the DET column of ol\_d\_id, and the ORE column of ol\_o\_id''. The ``stock'' table builds a joint index for ``DET column of s\_w\_id, DET column of s\_i\_id''. Here $\%d$ is a randomly generated value.

TPC-C supports users to customize tables of different sizes and we perform this group of experiments over six tables with different capacities. The ORE ciphertext column that builds the index belongs to the ``order\_line'' table. \autoref{table:TPCC——table} lists the different sizes of tables involved in the experiment and the number of tuples corresponding to the ``order\_line'' table.

\begin{table}[t]
    \centering
    \caption{Experimental Data}
    \begin{tabular}{|c|c|c|c|c|c|c|}
    \hline
        scale & 1 & 2 & 3 & 4& 5 &6 \\ \hline
        capacity (GB) & 5 & 10 & 15 & 20 & 25 &30  \\ \hline
        \#tuples in ``order\_line" ($\times 10^4$) & 30 & 60 & 90 & 120 & 150 &180 \\ \hline
    \end{tabular}
    \label{table:TPCC——table}
\end{table}
\par
In the scenario of read/write balanced, EncDB+IDX performs significantly better than Symmetria. Although write operation involves the time-consuming task for creating and updating indexes, experiments show that write operation for large tables does not change the performance advantage of EncDB+IDX. As shown in Fig.~\ref{fig:TPC-C1} and Fig.~\ref{fig:TPC-C2}, TPS and QPS of EncDB+IDX are 2.2 times that of Symmetria. Fig.~\ref{fig:TPC-C5} shows the latency in the read/write balanced scenario. The average latency of EncDB+IDX is only 20\% - 60\% of that of Symmetria. The throughput advantages of EncDB+IDX and postgresql+IDX, which maintain the index structure, do not diminish with the increase of table capacity, showing the excellent performance of reading and writing on large tables.

In the read-only scenario, as shown in Fig.~\ref{fig:TPC-C4} and Fig.~\ref{fig:TPC-C3}, the TPS and QPS of EncDB+IDX are about 45-160 times (1-2 orders of magnitude) that of Symmetria. The advantage of EncDB+IDX in read-only scenarios does not decrease with the increase of table volume. Fig.~\ref{fig:TPC-C6} shows that the latency of EncDB+IDX is the minimal and more stable than that of Symmetria, which reflects the important role of ciphertext index in query optimization. This advantage is constant as the number of rows in the table increases dramatically.


In addition, we also test the performance under high concurrency scenarios. We fix the volume of the table as scale=6 (according to Fig.~\ref{table:TPCC——table}), and vary the number of concurrent threads from 1 to 12. As shown in Fig.~\ref{fig:TPC-C1_thread}, Fig.~\ref{fig:TPC-C2_thread} and Fig.~\ref{fig:TPC-C5_thread}, in the scenario of read-write balance, the latency of the three solutions increase with the number of concurrent threads. When the number of threads is greater than 7, the TPS and QPS of EncDB+IDX and Symmetria both exhibit a fluctuating but downward trend. As the number of threads increases, PostgreSQL+IDX also suffers from performance degradation. When the number of thread is 12, the QPS of PostgreSQL+IDX decreases by 53\% compared with single thread scenario. The results implies us that the concurrency support of all solutions is limited.

\begin{figure}[t]
  \centering
  \includegraphics[width=.55\linewidth]{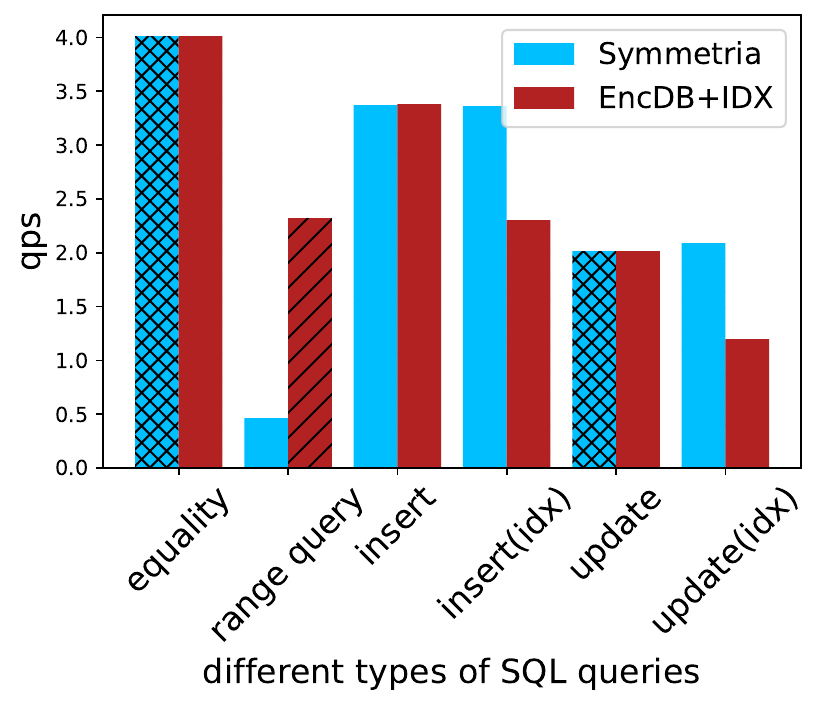}
  \vspace{0ex}
  \caption{Execution of different types of SQL statements}
  \label{different types of SQL}
  \vspace{0ex}
\end{figure}

Fig.~\ref{different types of SQL} shows the QPS that different types of SQL statements are executed separately in EncDB+IDX and Symmetria. ``$\mathsf{INSERT}$" and ``$\mathsf{UPDATE}$" indicate that the non-indexed columns are inserted and updated, and ``$\mathsf{INSERT}$ (IDX)" and ``$\mathsf{UPDATE}$ (IDX)" refers to the scheme that the indexed columns are inserted and updated. It can be seen that the two solutions have similar performance in write operation of the non-indexed ORE column. EncDB+IDX reduces the efficiency of ``$\mathsf{INSERT}$" and ``$\mathsf{UPDATE}$" operations on indexed columns, and is over two orders of magnitude better than Symmetrial in range queries.

\subsection{Security study}
\begin{table}[ht]
    \centering
    \caption{Encryption schemes used by EncDB, example SQL operations they allow over encrypted data on the server, and information revealed by each scheme's ciphertexts in the absence of any queries.}
    \label{security}
    \begin{tabular}{@{}p{4cm}p{4cm}p{4cm}@{}}
    \toprule
    Encryption scheme & SQL operations & Leakage\\
    \midrule
        Deterministic AES  & \(a = \text{const}, \text{GROUP BY}\) & Duplicates \\ 
        SAHE & \text{ADD},\text{ADP},\text{SUB} & Number of rows used to generate the result\\
        SMHE & \text{MUL},\text{MLP},\text{DIV} & Number of rows used to generate the result\\
        ORE & \( a >\text{const}, \text{MAX}\) & Order \\
    \bottomrule
    \end{tabular}
  \end{table}
  
\begin{table}[ht]
    \centering
    \caption{Number of distinct columns in the Sysbench-TPCC tables encrypted by EncDB under each of the encryption schemes shown. For each column, we consider only the weakest encryption scheme used. }
    \label{weakest}
    \begin{tabular}{@{}p{2cm}p{3cm}p{2cm}p{2cm}p{1cm}@{}}
    \toprule
    Table & Total columns & ORE & HE & DET \\
    \midrule
        warehouse  & 9 & 9&0&0\\ 
        district & 11 & 11&0&0\\
        customer & 21 & 21 &0&0\\
        history & 8 & 8&0&0 \\
        orders & 9&9&0&0 \\
        new orders & 3 & 3 & 0 & 0\\
        order line & 10 & 10 & 0 & 0\\
        stock & 17 &17&0&0\\
        item & 5 &5&0&0\\
    \bottomrule
    \end{tabular}
  \end{table}
Our software-only mode (with cipher index) guarantees IND-OCPA, and the other modes are at least as secure as the software-only mode. To understand the level of security that EncDB provides, it is important to consider the encryption schemes chosen by EncDB, since they leak different amounts of informations, as shown in Table \ref{security}. The worest is ORE, which reveals order, followed by DET. Table \ref{weakest} shows the encryption schemes that EncDB chooses for the Sysbench-TPCC workload. EncDB never reveals plaintext to the server. The weakest encryption scheme used, ORE, is always used.

The study by Naveed et al. \cite{Naveed_Kamara_Wright_2015}demonstrates that databases encrypted with Order Preserving Encryption (OPE) are highly susceptible to "inference attacks". Inference attacks aim to recover information about the data or queries by combining leaked information with publicly available data, such as census data or linguistic statistics. The most notable example of an inference attack is frequency analysis, used to crack classical ciphers. Muhammad et al. \cite{Naveed_Kamara_Wright_2015} conducted inference attacks on databases deploying attribute-preserving encryption schemes. Experiments showed that plaintext can be recovered from DET and OPE encrypted database columns using only the encrypted columns and publicly available auxiliary information.

EncDB utilizes the design scheme proposed by Kevin et al. \cite{boneh2015semantically}, which revisits the primitives related to Order-Revealing Encryption (ORE) schemes, redesigning them entirely based on symmetric encryption, hereinafter referred to as ExtendORE. It has been proven that the range query protocol constructed according to ExtendORE is robust against inference attacks proposed by Muhammad et al., and is more secure than all existing practical (stateless and non-interactive) OPE and ORE schemes, as well as being faster: encrypting a 32-bit integer requires only 55 microseconds, which is more than 65 times faster than the existing OPE schemes.

The security guarantees of an ORE-based encryption scheme assume that ORE encryption is computationally indistinguishable from random values with the same order pattern. After augmenting EncDB with an indexing mechanism, if two sequences exhibit the same order relation, an adversary cannot differentiate between the encryptions of the two value sequences.

Consider any adversary \textit{Adv} and any value sequences they request in the security game: \( \mathbf{v} = (v_1, \ldots, v_n) \) and \( \mathbf{w} = (w_1, \ldots, w_n) \). The two plaintext sequences have the same order relation. After the client-side encryption of \( \mathbf{v}, \mathbf{w} \), the adversary can observe the information received by the server in the range query protocol.

Next, we prove that the information learned by the adversary from EncDB and EncDB+IDX is theoretically the same.

Let us examine the information learned by the adversary from EncDB and EncDB+IDX when encrypting \( \mathbf{v}, \mathbf{w} \) on the client side.

Since ORE encryption is nondeterministic, the ciphertexts for the same plaintexts are also different. For EncDB, the adversary's retrieval operation is a full table scan, obtaining the search results through item-by-item comparison (using the \( \text{CMP} \) function). For two ciphertext sequences, the adversary can only know the order relation of each ciphertext sequence but cannot determine which ciphertexts belong to \( \mathbf{v} \) or \( \mathbf{w} \). For EncDB+IDX, since \( \mathbf{v} \) and \( \mathbf{w} \) have the same order relation, the update (or lookup) path along the tree split (insertion or probing) must be the same, and for the plaintexts in the plaintext sequences, the adversary cannot distinguish whether their plaintexts are in \( \mathbf{v} \) or \( \mathbf{w} \). Essentially, EncDB+IDX is equivalent to maintaining the entire EncDB table in B-tree in advance. It relies on operators based on the ORE\_EN type for comparison, and the operation logic of the operator is \( \text{CMP} \).

Therefore, it can be affirmed that the adversary facing EncDB and EncDB+IDX obtains information that is entirely consistent (only able to detect the order relation of plaintext). The "ideal" security of EncDB+IDX is IND-OCPA.

\vspace{-2ex}\section{Conclusion}\label{sec6}
In this work, we present and implement Enc$^2$DB, which is a encrypted database that is compatible with PostgreSQL and openGauss. On one hand, Enc$^2$DB implements a ciphertext index via UDT and UDO that is easily configured and natively supported by the query optimizer of both PostgreSQL and openGauss. On the other hand, Enc$^2$DB can be deployed in either software-only mode or TEE-enabled mode, each corresponds to different practical scenarios. In addition, in TEE-enabled mode we, for the first time, present a self-adaptive mode switch strategy that dynamically choose the suitable mode to execute a given query. The switch strategy fully utilizes the benefit of both cryptographic schemes and TEE at runtime.

\section*{Acknowledgment}
This work is partially supported by National Natural Science Foundation of China (No.61972309, 62272369), ZTE Industry-University-Institute Cooperation Funds (No. IA20230625001), and China 111 project.

%
%

\bibliographystyle{splncs04}
\bibliography{Enc2DB-DASFAA24}

\begin{thebibliography}{10}
\providecommand{\url}[1]{\texttt{#1}}
\providecommand{\urlprefix}{URL }
\providecommand{\doi}[1]{https://doi.org/#1}

\bibitem{ARM-TrustZone}
Arm trustzone (2009), \url{infocenter.arm.com/help/topic/com.arm.doc.prd29-genc-009492c/PRD29-GENC-009492C_trustzone_security_ whitepaper.pdf}

\bibitem{SGE-SDK}
Intel(r) software guard extensions sdk for linux* os. (2018), \url{https://download.01.org/intel-sgx/linux-2.2/docs/Intel_SGX_Developer_Reference_Linux_2.2_Open_Source.pdf}

\bibitem{intel-sgx}
Intel (2023), \url{https://www.intel.com/content/www/us/en/architecture-and-technology/software-guard-extensions.html}

\bibitem{agrawal2004order}
Agrawal, R., Kiernan, J., Srikant, R., Xu, Y.: Order preserving encryption for numeric data. In: Proceedings of the 2004 ACM SIGMOD international conference on Management of data. pp. 563--574 (2004)

\bibitem{antonopoulos2020azure}
Antonopoulos, P., Arasu, A., Singh, K.D., Eguro, K., Gupta, N., Jain, R., Kaushik, R., Kodavalla, H., Kossmann, D., Ogg, N., et~al.: Azure sql database always encrypted. In: Proceedings of the 2020 ACM SIGMOD International Conference on Management of Data. pp. 1511--1525 (2020)

\bibitem{ArvindArasu2013OrthogonalSW}
Arasu, A., Blanas, S., Eguro, K., Kaushik, R., Kossmann, D., Ramamurthy, R., Venkatesan, R.: Orthogonal security with cipherbase. conference on innovative data systems research  (2013)

\bibitem{arnautov2016scone}
Arnautov, S., Trach, B., Gregor, F., Knauth, T., Martin, A., Priebe, C., Lind, J., Muthukumaran, D., O'keeffe, D., Stillwell, M.L., et~al.: Scone: Secure linux containers with intel sgx. In: 12th USENIX Symposium on Operating Systems Design and Implementation (OSDI 16). pp. 689--703 (2016)

\bibitem{bailleu2019speicher}
Bailleu, M., Thalheim, J., Bhatotia, P., Fetzer, C., Honda, M., Vaswani, K.: Speicher: Securing lsm-based key-value stores using shielded execution. In: 17th USENIX Conference on File and Storage Technologies (FAST 19). pp. 173--190 (2019)

\bibitem{boldyreva2011order}
Boldyreva, A., Chenette, N., O’Neill, A.: Order-preserving encryption revisited: Improved security analysis and alternative solutions. In: Annual Cryptology Conference. pp. 578--595. Springer (2011)

\bibitem{boneh2015semantically}
Boneh, D., Lewi, K., Raykova, M., Sahai, A., Zhandry, M., Zimmerman, J.: Semantically secure order-revealing encryption: Multi-input functional encryption without obfuscation. In: Annual International Conference on the Theory and Applications of Cryptographic Techniques. pp. 563--594. Springer (2015)

\bibitem{chenette2016practical}
Chenette, N., Lewi, K., Weis, S.A., Wu, D.J.: Practical order-revealing encryption with limited leakage. In: International conference on fast software encryption. pp. 474--493. Springer (2016)

\bibitem{1555162}
Conti, M., Di~Pietro, R., Mancini, L.V., Mei, A.: (old) distributed data source verification in wireless sensor networks. Inf. Fusion  \textbf{10}(4),  342--353 (2009). \doi{http://dx.doi.org/10.1016/j.inffus.2009.01.002}

\bibitem{costan2016intel}
Costan, V., Devadas, S.: Intel sgx explained. Cryptology ePrint Archive  (2016)

\bibitem{damgaard2001generalisation}
Damg{\aa}rd, I., Jurik, M.: A generalisation, a simplification and some applications of paillier's probabilistic public-key system. In: International workshop on public key cryptography. pp. 119--136. Springer (2001)

\bibitem{elgamal1985public}
ElGamal, T.: A public key cryptosystem and a signature scheme based on discrete logarithms. IEEE transactions on information theory  \textbf{31}(4),  469--472 (1985)

\bibitem{eskandarian2017oblidb}
Eskandarian, S., Zaharia, M.: Oblidb: Oblivious query processing for secure databases. arXiv preprint arXiv:1710.00458  (2017)

\bibitem{fousse2011benaloh}
Fousse, L., Lafourcade, P., Alnuaimi, M.: Benaloh’s dense probabilistic encryption revisited. In: International Conference on Cryptology in Africa. pp. 348--362. Springer (2011)

\bibitem{gentry2009fully}
Gentry, C.: Fully homomorphic encryption using ideal lattices. In: Proceedings of the forty-first annual ACM symposium on Theory of computing. pp. 169--178 (2009)

\bibitem{hacigumucs2002executing}
Hacig{\"u}m{\"u}{\c{s}}, H., Iyer, B., Li, C., Mehrotra, S.: Executing sql over encrypted data in the database-service-provider model. In: Proceedings of the 2002 ACM SIGMOD international conference on Management of data. pp. 216--227 (2002)

\bibitem{AMD-SEV}
Kaplan, D., Powell, J., Woller, T.: Amd sev, \url{http://amd-dev.wpengine.netdna-cdn.com/wordpress/media/2013/12/AMD_Memory_Encryption_Whitepaper_v7-Public.pdf}

\bibitem{kim2019shieldstore}
Kim, T., Park, J., Woo, J., Jeon, S., Huh, J.: Shieldstore: Shielded in-memory key-value storage with sgx. In: Proceedings of the Fourteenth EuroSys Conference 2019. pp. 1--15 (2019)

\bibitem{kopytov2017sysbench}
Kopytov, A.: Sysbench. \url{https://github.com/akopytov/sysbench} (2017)

\bibitem{cryptoeprint:2010:264}
Krawczyk, H.: Cryptographic extraction and key derivation: The hkdf scheme. Cryptology ePrint Archive, Report 2010/264 (2010)

\bibitem{mishra2018oblix}
Mishra, P., Poddar, R., Chen, J., Chiesa, A., Popa, R.A.: Oblix: An efficient oblivious search index. In: 2018 IEEE Symposium on Security and Privacy (SP). pp. 279--296. IEEE (2018)

\bibitem{naccache1998new}
Naccache, D., Stern, J.: A new public key cryptosystem based on higher residues. In: Proceedings of the 5th ACM Conference on Computer and Communications Security. pp. 59--66 (1998)

\bibitem{Naveed_Kamara_Wright_2015}
Naveed, M., Kamara, S., Wright, C.V.: Inference attacks on property-preserving encrypted databases. In: Proceedings of the 22nd ACM SIGSAC Conference on Computer and Communications Security (2015)

\bibitem{okamoto1998epoc}
Okamoto, T., Uchiyama, S., Fujisaki, E.: Epoc: Efficient probabilistic public-key encryption (submission to p1363a). IEEE P1363a p.~18 (1998)

\bibitem{orenbach2017eleos}
Orenbach, M., Lifshits, P., Minkin, M., Silberstein, M.: Eleos: Exitless os services for sgx enclaves. In: Proceedings of the Twelfth European Conference on Computer Systems. pp. 238--253 (2017)

\bibitem{paillier1999public}
Paillier, P.: Public-key cryptosystems based on composite degree residuosity classes. In: International conference on the theory and applications of cryptographic techniques. pp. 223--238. Springer (1999)

\bibitem{pandey2012property}
Pandey, O., Rouselakis, Y.: Property preserving symmetric encryption. In: Annual International Conference on the Theory and Applications of Cryptographic Techniques. pp. 375--391. Springer (2012)

\bibitem{papadimitriou2016big}
Papadimitriou, A., Bhagwan, R., Chandran, N., Ramjee, R., Haeberlen, A., Singh, H., Modi, A., Badrinarayanan, S.: Big data analytics over encrypted datasets with seabed. In: 12th USENIX symposium on operating systems design and implementation (OSDI 16). pp. 587--602 (2016)

\bibitem{poddar2016arx}
Poddar, R., Boelter, T., Popa, R.A.: Arx: A strongly encrypted database system. IACR Cryptol. ePrint Arch.  \textbf{2016}, ~591 (2016)

\bibitem{popa2012cryptdb}
Popa, R.A., Redfield, C.M., Zeldovich, N., Balakrishnan, H.: Cryptdb: processing queries on an encrypted database. Communications of the ACM  \textbf{55}(9),  103--111 (2012)

\bibitem{priebe2018enclavedb}
Priebe, C., Vaswani, K., Costa, M.: Enclavedb: A secure database using sgx. In: 2018 IEEE Symposium on Security and Privacy (SP). pp. 264--278. IEEE (2018)

\bibitem{rivest1978data}
Rivest, R.L., Adleman, L., Dertouzos, M.L., et~al.: On data banks and privacy homomorphisms. Foundations of secure computation  \textbf{4}(11),  169--180 (1978)

\bibitem{rivest1978method}
Rivest, R.L., Shamir, A., Adleman, L.: A method for obtaining digital signatures and public-key cryptosystems. Communications of the ACM  \textbf{21}(2),  120--126 (1978)

\bibitem{sabt2015trusted}
Sabt, M., Achemlal, M., Bouabdallah, A.: Trusted execution environment: what it is, and what it is not. In: 2015 IEEE Trustcom/BigDataSE/ISPA. vol.~1, pp. 57--64. IEEE (2015)

\bibitem{savvides2020efficient}
Savvides, S., Khandelwal, D., Eugster, P.: Efficient confidentiality-preserving data analytics over symmetrically encrypted datasets. Proceedings of the VLDB Endowment  \textbf{13}(8),  1290--1303 (2020)

\bibitem{sun2021building}
Sun, Y., Wang, S., Li, H., Li, F.: Building enclave-native storage engines for practical encrypted databases. Proceedings of the VLDB Endowment  \textbf{14}(6),  1019--1032 (2021)

\bibitem{monomi}
Tu, S., Kaashoek, M.F., Madden, S., Zeldovich, N.: Processing analytical queries over encrypted data. very large data bases  (2013)

\bibitem{tpccsysbench}
Vadim, T., Alexey, S., Alexey, K., Sebastian, D.: sysbench-tpcc. \url{https://github.com/Percona-Lab/sysbench-tpcc/} (2018)

\bibitem{vinayagamurthy2019stealthdb}
Vinayagamurthy, D., Gribov, A., Gorbunov, S.: Stealthdb: a scalable encrypted database with full sql query support. Proc. Priv. Enhancing Technol.  \textbf{2019}(3),  370--388 (2019)

\bibitem{sdb}
Wong, W.K., Kao, B., Cheung, D.W.L., Li, R., Yiu, S.M.: Secure query processing with data interoperability in a cloud database environment. In: Proceedings of the 2014 ACM SIGMOD international conference on Management of data. pp. 1395--1406 (2014)

\bibitem{DBLP:journals/pvldb/XiaZZZCEFFKF22}
Xia, S., Zhu, Z., Zhu, C., Zhao, J., Chard, K., Elmore, A.J., Foster, I.T., Franklin, M.J., Krishnan, S., Fernandez, R.C.: Data station: Delegated, trustworthy, and auditable computation to enable data-sharing consortia with a data escrow. Proceedings of the VLDB Endowment  \textbf{15}(11),  3172--3185 (2022)

\bibitem{zheng2017opaque}
Zheng, W., Dave, A., Beekman, J.G., Popa, R.A., Gonzalez, J.E., Stoica, I.: Opaque: An oblivious and encrypted distributed analytics platform. In: 14th USENIX Symposium on Networked Systems Design and Implementation (NSDI 17). pp. 283--298 (2017)

\bibitem{zhu2021full}
Zhu, J., Cheng, K., Liu, J., Guo, L.: Full encryption: An end to end encryption mechanism in gaussdb. Proceedings of the VLDB Endowment  \textbf{14}(12),  2811--2814 (2021)

\end{thebibliography}

\end{document}